\documentclass[prd,showpacs,superscriptaddress,nofootinbib,floatfix,11pt]{revtex4-2}

\usepackage[utf8]{inputenc}
\usepackage[T1]{fontenc}
\usepackage{textgreek}
\usepackage{amsmath,amssymb}
\usepackage{slashed}
\usepackage{subfigure}
\usepackage[colorlinks=true,
breaklinks=true,
urlcolor=magenta,
citecolor=blue]{hyperref}
\usepackage[usenames,dvipsnames]{color}
\usepackage{appendix}
\usepackage{braket,bm}
\usepackage{multirow}
\usepackage{enumitem}
\usepackage{color}
\usepackage{slashed}
\usepackage{orcidlink}
\usepackage{graphicx}
\usepackage{tikz}
\usepackage{tikz-feynman}
\usepackage{footnotebackref}
\usepackage{booktabs}
\usepackage{array}
\usepackage{overpic}
\usepackage{float}
\usepackage{threeparttable}
\usepackage{bbm}  
\usepackage{mathrsfs}
\allowdisplaybreaks[4]

\newcommand{\re}{\operatorname{Re}}
\newcommand{\im}{\operatorname{Im}}
\newcommand{\mex}{{\mathbbm{m}}}
\newcommand{\nch}{N}
\newcommand{\Log}{\operatorname{Log}}

\newcommand{\uestc}{\affiliation{School of Physics, University of Electronic Science and Technology of China, Chengdu 611731, China}} 

\newcommand{\itp}{\affiliation{Institute of Theoretical Physics, Chinese Academy of Sciences, Beijing 100190, China}}

\newcommand{\ucas}{\affiliation{School of Physical Sciences, University of Chinese Academy of Sciences, Beijing 100049, China}}

\newcommand{\scnt}{\affiliation{Southern Center for Nuclear-Science Theory (SCNT), Institute of Modern Physics, Chinese Academy of Sciences, Huizhou 516000, China}}

\newcommand{\sxu}{\affiliation{College of Physics and Electronic Engineering, Shanxi University, Taiyuan 030006, China}}

\begin{document}
\title{Coupled-channel extension of the effective-range expansion with left-hand cut}
\author{Bing Wu\orcidlink{0009-0004-8178-3015}}\email{wu.bing@uestc.edu.cn}
\uestc 

\author{Wen-Jia Wang\orcidlink{0000-0001-5092-0583}}\email{wwenjia@std.uestc.edu.cn}
\uestc

\author{Meng-Lin Du\orcidlink{0000-0002-7504-3107}}\email{ du.ml@uestc.edu.cn}
\uestc\scnt

\author{Feng-Kun Guo\orcidlink{0000-0002-2919-2064}}\email{fkguo@itp.ac.cn}
\itp \ucas \scnt

\author{Hao-Jie Jing}\email{jinghaojie@sxu.edu.cn}
\sxu

\begin{abstract}
We generalize the recently proposed effective-range expansion with a left-hand cut [\href{https://link.aps.org/doi/10.1103/cwdt-dj6z}{Phys. Rev. Lett. 135, 011903 (2025)}] to a coupled-channel framework. By systematically analyzing the singularity structure of the $S$-wave coupled-channel one-particle exchange (OPE) potential under various mass configurations, we derive an analytical parametrization of the coupled-channel scattering amplitude that incorporates the OPE left-hand cut for multiple channels. Taking the two-channel case as an illustration, we benchmark this parametrization against the Lippmann--Schwinger equation (LSE) by comparing the amplitudes and their inverses at different expansion orders. Even at the lowest order, with only six parameters, the coupled-channel parametrization faithfully reproduces the LSE results and enables precise extraction of the coupling constants, demonstrating its reliability for two-body coupled-channel scattering problems involving particle exchange.
\end{abstract}

\maketitle
\newpage

\section{Introduction}
The effective range expansion (ERE) \cite{Bethe:1949yr,Blatt:1949zz} is widely employed in the analysis of near-threshold phenomena. It offers a model-independent parametrization of the low-energy two-body scattering amplitude based on unitarity and analyticity. The relevance of this framework has grown considerably with the discovery of numerous exotic hadronic states and candidates lying close to two-body thresholds.
For two-particle $S$-wave elastic scattering in nonrelativistic kinematics, the unitarized amplitude takes the form
\begin{align}
    T(k)=-\frac{2\pi}{\mu}\frac{1}{k\cot\delta-ik}~,\label{equ:unitaryofT}
\end{align} 
where $\mu$ is the reduced mass of the two-body system, $k$ is the magnitude of the center-of-mass (c.m.) momentum, and \( \delta \) is the phase shift. The conventional ERE expands $k\cot\delta$ as a power series in $k^2$ in the near-threshold region:
\begin{align}
k \cot \delta = \frac{1}{a} + \frac{1}{2} r k^2 + \mathcal{O}(k^4),\label{equ:tradition_ERE}
\end{align}
where $a$ and $r$ are the scattering length and effective range, respectively. This parameterization~(\ref{equ:unitaryofT},~\ref{equ:tradition_ERE}) can also be obtained by summing the geometric series of Feynman diagrams for contact interactions~\cite{Kaplan:1998we,Braaten:2005jj}. 
Within the framework for coupled channels, the nonrelativistic $S$-wave coupled-channel inverse amplitude can be parameterized at the lowest order as (illustrated here for the two-channel case)~\cite{Ross:1960mert,Ross:1961jlg,Cohen:2004kf,Braaten:2005jj,Hanhart:2011jz,CLQCD:2019npr,Dong:2020hxe} 
\begin{align}
    T^{-1}=\begin{pmatrix}
       -\frac{1}{a_{11}}+i\frac{\mu_1k_1}{2\pi} & \frac{1}{a_{12}} \\
       \frac{1}{a_{12}} & -\frac{1}{a_{22}}+i\frac{\mu_2k_2}{2\pi}
      \end{pmatrix},\label{equ:CCERE_traditional}
\end{align}
where $k_i$ ($i = 1,2$) denotes the magnitude of the c.m. momentum for channel $i$, and $a_{ij}$ ($i,j = 1,2$) are scattering parameters.\footnote{The general parameterizations of the coupled-channel amplitude can be expressed in terms of scattering phase shifts and inelastic coefficients; see e.g., in Refs.~\cite{Blatt:1952zz,Stapp:1956mz,Hoferichter:2012wf,Guo:2013rpa}.} 

While the ERE provides an excellent low-energy description for short-range interactions, the radius of its convergence domain is set by the singularity closest to the threshold, and can be severely restricted in the presence of long-range forces. In particular, a left-hand cut (lhc) emerges near threshold when a relatively light particle is exchanged in the $t$-channel, or when the $u$-channel exchange involves a particle whose mass lies just above the mass difference of the two scattering particles~\cite{Du:2024gzw,Du:2025vkm,Wang:2026ups,Liu:2026xrk}. The lhc branch point can be the singularity that sets the radius of convergence of the traditional ERE. 
Furthermore, one-particle exchange (OPE) may also generate, through its interplay with short-range interactions, a zero of the $S$-wave amplitude, equivalently a pole of $k\cot\delta$, as demonstrated, e.g., for $DD^*$ scattering~\cite{Du:2023hlu} and the scattering between a neutron and a one-neutron-halo nucleus~\cite{ Zhang:2023wdz}. Depending on the interaction strengths, this pole may lie closer to threshold than the nearest lhc branch point and thus further shrink the domain of convergence.
To systematically incorporate such OPE-induced singularities, a generalized ERE for elastic scattering that explicitly includes the lhc in a closed form was recently proposed in Ref.~\cite{Du:2024gzw}, with higher-order improvements discussed in Ref.~\cite{Wang:2026ups}. Instead of Eq.~\eqref{equ:tradition_ERE}, one writes
 \begin{align}
k\cot{\delta}&=\frac{\tilde{d}(k^2)- \tilde{g} d^\text{R}(k^2)}{\tilde{n}(k^2) + \tilde{g}\left[ L(k^2)-L(0)\right]}\ ,\label{eq:modfere}
\end{align}
where $\tilde{d}(k^2)$ and $\tilde{n}(k^2)$, which is normalized to $\tilde{n}(0)=1$, are polynomials (or, more generally, rational functions) in $k^2$. The logarithmic functions $d^\text{R}(k^2)$ and $L(k^2)$, whose expressions can be found in Refs.~\cite{Du:2024gzw,Wang:2026ups}, capture the analytic structure introduced by the lhc. The parameter $\tilde{g}$ is related to the coupling to the exchanged particle, thereby controlling the strength of the lhc. In the absence of the lhc ($\tilde{g}=0$), setting $\tilde{n}(k^2)=1$ reduces Eq.~\eqref{eq:modfere} to the traditional ERE of Eq.~\eqref{equ:tradition_ERE}. 
The amplitude zero mentioned above is also naturally incorporated. 
This framework has been successfully applied to reproduce $DD^*$ scattering amplitudes obtained from the Lippmann-Schwinger equation (LSE) for a potential including contact terms and one-pion exchange~\cite{Du:2024gzw}. The framework has also been applied in analyzing the nucleon-nucleon scattering phase shifts~\cite{Liu:2026xrk}; the amplitude zero for the center-of-mass momentum about $0.35$ GeV in the $^1S_0$ $pn$ partial wave is naturally accommodated, and the axial pion-nucleon coupling constant can be extracted, though with a sizable uncertainty.
One notices that finite-range one-particle exchange interactions can also be treated with distorted-wave theory~\cite{Nieves:2003uu, Entem:2025siq}.
For comprehensive discussions of singularities due to on-shell particles, see Refs.~\cite{Cutkosky:1960sp,Eden:1966dnq,Jing:2025qmi}; kinematic singularities are addressed in, e.g., Refs.~\cite{Martin:1970hmp,Wu:2023uva,Wu:2024xwy}.

The above discussion highlights several advantages of the generalized ERE in Eq.~\eqref{eq:modfere} over the traditional form in Eq.~\eqref{equ:tradition_ERE}: First, it explicitly incorporates the near-threshold OPE lhc, thereby extending the domain of convergence. Second, it inherently accounts for the amplitude zero, i.e., poles in $k\cot\delta$. Third, it enables a model-independent extraction of the coupling strength of the scattered particles to the exchanged particle, directly from the experimental data or lattice quantum chromodynamics (QCD) results. 
The generalized ERE given by Eq.~\eqref{eq:modfere} applies only to single-channel scatterings. 
Thus, its applicability regime can be limited by the presence of the next threshold, which brings about a square-root branch point in $k^2$.
Yet many resonances lie close to more than one threshold of hadron pairs to which they can couple.
A prominent example is given by the $\chi_{c1}(3872)$ (also known as $X(3872)$) whose mass is extremely close to the $D^0\bar{D}^{*0}$ threshold~\cite{ParticleDataGroup:2026mpi},\footnote{From a combined analysis of the LHCb~\cite{LHCb:2020fvo,LHCb:2022jez} and BESIII~\cite{BESIII:2023hml} data, the $X(3872)$ mass has been determined to be $160_{-74}^{+57}$~keV below the $D^{0}\bar D^{*0}$ threshold~\cite{Ji:2025hjw}.} while the $D^+D^{*-}$ threshold is only $8.2$ MeV higher.
Most of these near-threshold coupled-channel effects originate from approximate symmetry derived from QCD, such as isospin symmetry, SU(3) flavor symmetry, and heavy quark spin symmetry. A single-channel parameterization in Eq.~\eqref{eq:modfere} alone is simply not sufficient. The present work therefore aims to extend this framework to the coupled-channel case, i.e., to generalize Eq.~\eqref{equ:CCERE_traditional} to incorporate near-threshold OPE lhcs.

In this work, we systematically analyze the singularity structure of the $S$-wave coupled-channel OPE potential under various mass configurations. Using dispersion relations and the $N/D$ method, we derive an analytical parametrization of the coupled-channel scattering amplitude that incorporates the OPE lhcs for multiple channels. To validate our approach, we explicitly compare the amplitude and its inverse with the LSE solution for a two-channel case. By increasing the expansion order, we further demonstrate the reliability and stability of the proposed scheme. 

The paper is organized as follows. In Sec.~\ref{sec:formalism}, we present the theoretical framework, including a systematic analysis of the singularity structure of the $S$-wave coupled-channel OPE potential and corresponding unitarized amplitude, the $N/D$ method, dispersion relations, and the explicit parametrization for the two-channel case. In Sec.~\ref{sec:results}, we compare the amplitude and its inverse obtained from the coupled-channel parametrization with the corresponding LSE solutions for a two-channel case. A brief summary is given in Sec.~\ref{sec:summary}. Various technical and pedagogical details are relegated to the appendices.

\section{Formalism}\label{sec:formalism}
\begin{figure}[tb]
    \centering
    \begin{tikzpicture}[
        line width=0.8pt, 
        scale=1.5,
        midarrow/.style={
            postaction={decorate},
            decoration={
                markings,
                mark=at position 0.6 with {\arrow{latex}}
            }
        }
        ]
        
        \draw[midarrow] (0,0.25) node[left] {$m_{i1}$} -- (1,0);
        \draw[midarrow] (1,0) -- (2,0.25) node[right] {$m_{j1}$};
        \draw[midarrow] (0,-1.25) node[left] {$m_{i2}$} -- (1,-1);
        \draw[midarrow] (1,-1) -- (2,-1.25) node[right] {$m_{j2}$};
        \draw[dashed] (1,0) -- node[left] {$\mex_{t,ij}$} (1,-1);

        \draw[midarrow] (4,0.25) node[left] {$m_{i1}$} -- (5,0);
        \draw[midarrow] (5,0) -- (6,-1.25) node[right] {$m_{j2}$};
        \draw[midarrow] (4,-1.25) node[left] {$m_{i2}$} -- (5,-1);
        \draw[midarrow] (5,-1) -- (6,0.25) node[right] {$m_{j1}$};
        \draw[dashed] (5,0) -- node[left] {$\mex_{u,ij}$} (5,-1);
        \node at (1,-1.5) {$(a)$};
        \node at (5,-1.5) {$(b)$};
        
    \end{tikzpicture} 
    \caption{One-particle exchange Feynman diagrams for (a) $t$-channel and (b) $u$-channel scattering processes. External particle masses are denoted by $m_{\alpha l}$, with channel index $\alpha=i,j$ and particle index $l=1,2$. The masses of the exchanged particles in the $t$- and $u$-channels are $\mex_{t,ij}$ and $\mex_{u,ij}$, respectively.}
    \label{fig:feyndiag}
\end{figure}
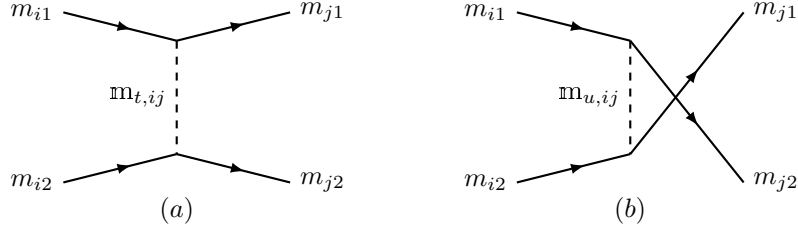

For an $S$-wave scattering, as depicted in Fig.~\ref{fig:feyndiag}, the lhc contributions from the $t$- and $u$-channel exchange partial waves are given by
\begin{align}
    L_{t,ij}(k_j^2)&=\frac{1}{2}\int_{-1}^{1}\frac{{\rm d}\cos\theta}{t_{ij}-\mex_{t,ij}^2}\notag\\
    &=\frac{1}{4k_ik_j}\log\left[ \frac{m_{i1}^2+m_{j1}^2-2\sqrt{k_i^2+m_{i1}^2}\sqrt{k_j^2+m_{j1}^2}-\mex_{t,ij}^2+2k_i k_j}{m_{i1}^2+m_{j1}^2-2\sqrt{k_i^2+m_{i1}^2}\sqrt{k_j^2+m_{j1}^2}-\mex_{t,ij}^2-2k_i k_j} \right]\ ,\label{Lt_origin}\\
    L_{u,ij}(k_j^2)&=\frac{1}{2}\int_{-1}^{1}\frac{{\rm d}\cos\theta}{u_{ij}-\mex_{u,ij}^2}\notag\\ 
    &=\frac{1}{4k_ik_j}\log\left[ \frac{m_{i1}^2+m_{j2}^2-2\sqrt{k_i^2+m_{i1}^2}\sqrt{k_j^2+m_{j2}^2}-\mex_{u,ij}^2+2k_i k_j}{m_{i1}^2+m_{j2}^2-2\sqrt{k_i^2+m_{i1}^2}\sqrt{k_j^2+m_{j2}^2}-\mex_{u,ij}^2-2k_i k_j} \right]\ ,\label{Lu_origin}
\end{align}
respectively. Here, the Mandelstam variables are defined as
\begin{align}
    t_{ij}&=m_{i1}^2 + m_{j1}^2 -2\sqrt{k_i^2+m_{i1}^2}\sqrt{k_j^2 + m_{j1}^2} + 2k_ik_j\cos\theta\ ,\label{defoft}\\
    u_{ij}&=m_{i1}^2 + m_{j2}^2 -2\sqrt{k_i^2+m_{i1}^2}\sqrt{k_j^2 + m_{j2}^2} - 2k_ik_j\cos\theta\ ,\label{defofu}
\end{align}
where $k_\alpha$ denotes the magnitude of the three-momentum in the $\alpha$-th channel, and $m_{\alpha l}$ represents the mass of a scattering particle, with channel index $\alpha=i,j$ and particle index $l=1,2$, $\theta$ is the scattering angle. The quantities $\mex_{t,ij}$ and $\mex_{u,ij}$ are the masses of the exchanged particles for the $t$- and $u$-channel, respectively. For $i=j$, Eqs.~(\ref{Lt_origin}) and (\ref{Lu_origin}) reduce to the single-channel results derived in Refs.~\cite{Du:2024gzw,Du:2025vkm,Wang:2026ups}. 
For $i\neq j$, $L_{u,ij}(k_j^2)$ can be obtained from $L_{t,ij}(k_j^2)$ by simply interchanging $m_{j1}$ and $m_{j2}$ and correspondingly replacing $\mex_{t,ij}$ with $\mex_{u,ij}$. Therefore, without loss of generality, in the subsequent analysis we will focus on $L_{t,ij}(k_j^2)$.  

Moreover, $k_i^2$ and $k_j^2$ are related through the energy conservation condition
\begin{align}
\sqrt{k_i^2+m_{i1}^2}+\sqrt{k_i^2+m_{i2}^2}=\sqrt{k_j^2+m_{j1}^2}+\sqrt{k_j^2+m_{j2}^2}\ ,\label{equ:energyconservation}
\end{align}
i.e., $k_i^2$ and $k_j^2$ are not independent. For the convenience of the subsequent analysis, we choose to work with $L_{t,ij}(k_j^2)$ rather than $L_{t,ij}(k_i^2)$. Furthermore, as this work does not address three-body scattering, we restrict the exchanged particle mass to 
\begin{align}
    \mex_{t,ij}&>{\rm max}\{|m_{i1}-m_{j1}|,|m_{i2}-m_{j2}|\}\ .\label{constr_three_t}
\end{align}
This condition ensures the existence of a near-threshold region within the physical kinematical domain that is free from on-shell three-body intermediate states. Within this region and in the immediate subthreshold region, it suffices to consider only the OPE lhc and the two-body unitarity cut.

\subsection{Near-threshold approximation for $L_{t,ij}(k_j^2)$}

We consider the situation such that the thresholds of the coupled channels are not far from each other, such that the nonrelativistic approximation is applicable to all channels in the near-threshold regime.
Then energy conservation condition in Eq.~(\ref{equ:energyconservation}) is expanded to the nonrelativistic form,
\begin{align}
    m_{i\rm{th}}+\frac{k_i^2}{2\mu_{i}}+\mathcal{O}(k_i^4)=m_{j\rm{th}}+\frac{k_j^2}{2\mu_{j}}+\mathcal{O}(k_j^4)\ ,\label{nonrelativisofEcons}
\end{align}
where $m_{\alpha\rm{th}}=m_{\alpha1}+m_{\alpha2}$ and $\mu_\alpha=m_{\alpha1}m_{\alpha2}/m_{\alpha\rm{th}}$ denote the threshold and reduced mass of the two-particle system in the $\alpha$-th channel, respectively. Retaining terms up to $\mathcal{O}(k_\alpha^2)$ in Eq.~(\ref{nonrelativisofEcons}) yields
\begin{align}
    k_j^2=\frac{\mu_j}{\mu_i}k_i^2 + 2\mu_j(m_{i\rm{th}}-m_{j\rm{th}})\ .\label{relationkikj}
\end{align} 
Substituting the expansion\footnote{Equation~(\ref{nonrelativisofEcons}) implies $m_{i\rm{th}}-m_{j\rm{th}}=\mathcal{O}(k_\alpha^2)$, so the term $k_i^2\mu_{j}(m_{i\rm{th}}-m_{j\rm{th}})/(2m_{i1}m_{j1})$ arising from $k_i^2k_j^2/(4m_{i1}m_{j1})$ is of higher order and can be dropped.}
\begin{align}
\sqrt{k_i^2+m_{i1}^2}\sqrt{k_j^2+m_{j1}^2}= m_{i1}m_{j1} + \frac{m_{j1}}{2m_{i1}}k_i^2 + \frac{m_{i1}}{2m_{j1}}k_j^2+\mathcal{O}(k_\alpha^4)\label{nonrelativisticofsqrt}
\end{align}
in Eq.~(\ref{defoft}), and neglecting all terms of $\mathcal{O}(k_\alpha^4)$ we arrive at
\begin{align}
    L_{t,ij}(k_j^2) = \frac{1}{4k_ik_j}\left[\log f_{-,ij}(k_j^2) -\log f_{+,ij}(k_j^2)\right], \label{equLt_nonreal}
\end{align}
where
\begin{align}
    f_{\pm,ij}(k_j^2)=\mex_{t,ij}^2-(m_{i1}-m_{j1})^2+\left( \sqrt{\frac{m_{j1}}{m_{i1}}}k_i\pm \sqrt{\frac{m_{i1}}{m_{j1}}}k_j\right)^2.
\end{align}

\subsection{Singularity structure of $L_{t,ij}(k_j^2)$}\label{singularityofL}

To facilitate the subsequent dispersion analysis, we first examine the singularity structure of $L_{t,ij}(k_j^2)$ on the complex $k_j^2$-plane under various mass configurations of the involved particles.

For the diagonal elements ($i=j$), the situation reduces to the single-channel case analyzed in Refs.~\cite{Du:2024gzw,Du:2025vkm,Wang:2026ups}. All singularities appear exclusively on the negative real axis and give rise to the lhc. In this work, we focus on $S$-wave scattering. Along the lhc, the discontinuities for the $t$- and $u$-channel OPE potentials are proportional to the corresponding imaginary parts given by
\begin{align}
    \im L_{t,ii}(k_i^2)&=-\frac{\pi}{4k_i^2}\ ,\quad \text{for } k_i^2<-\frac{\mex_{t,ii}^2}{4}\ ,\label{equ:ImLtii}\\
    \im L_{u,ii}(k_i^2)&=-\frac{\pi}{4k_i^2}\ ,\quad \text{for } -\frac{\mu_{+,i}^2}{4\eta_i^2}<k_i^2<-\frac{\mu_{+,i}^2}{4}\ ,\label{equ:ImLuii}
\end{align}
where $\mu_{+,i}^2 \equiv 4\mu_i\mu_{\text{ex},i}^2/m_{i\text{th}}$, $\eta_i \equiv |\Delta_i|/m_{i\text{th}}$, with $\mu_{\text{ex},i}^2 \equiv \mex_{u,ii}^2 - \Delta_i^2$, and $\Delta_i \equiv m_{i1} - m_{i2}$.

The off-diagonal elements ($i \neq j$) exhibit a more intricate singularity structure, which depends sensitively on the masses of the particles involved. For convenience, we label the channels in ascending order of their two-body thresholds such that
\begin{align}
m_{i\rm th} > m_{j\rm th}\ , \quad \text{for } i > j \ .
\end{align}
Time-reversal invariance guarantees $L_{t,ij} = L_{t,ji}$. A classification of the singularity structures in $L_{t,ij}(k_j^2)$ of various cases for $i>j$ is presented in Appendix~\ref{sec:appendixofsingularity}. From the linear relation between $k_i^2$ and $k_j^2$ in Eq.~\eqref{relationkikj}, it follows that if $k_i^2$ is used as the variable, the singularity structure of $L_{t,ij}(k_i^2)$ on the complex $k_i^2$-plane would simply be a globally shifted and scaled version of that of $L_{t,ij}(k_j^2)$ on the complex $k_j^2$-plane, as described in Appendix~\ref{sec:appendixofsingularity}. 

Here, we focus on the most typical scenario in Appendix~\ref{sec:appendixofsingularity}, characterized by the condition 
\begin{align}\label{equ:condition}
\mex_{t,ij}^2>(m_{i1}-m_{j1})^2 + 2(m_{j\rm th}-m_{i\rm th})\frac{m_{i1}}{m_{j1}}\mu_j\ ,& \quad \text{ for } i<j\ . 
\end{align}
In this case, the imaginary part of $L_{t,ij}(k_j^2)$ takes the simple form
\begin{align}
\operatorname{Im} L_{t,ij}(k_j^2) = -\frac{\pi}{4k_i k_j}, \qquad \text{for } k_j^2 \in [-M_{ji,-}^2, -M_{ji,+}^2],\label{equ:ImLij}
\end{align}
where the endpoints $-M_{ji,\pm}^2$ denote the roots of $f_{\pm,ij}(k_j^2)=0$, namely
\begin{align}
f_{\pm,ij}(k_j^2=-M_{ji,\pm}^2)=0\ .\label{equ:definationofMij}
\end{align}
The explicit analytic expression for $-M_{ji,\pm}^2$ is rather cumbersome and is omitted here. In practice, once the particle masses $m_{\alpha l}$ and the exchange mass $\mex_{t,ij}$ are specified, these endpoints can be readily obtained by numerically solving Eq.~(\ref{equ:definationofMij}). In particular, for $i = j$, Eq.~\eqref{equ:definationofMij} yields 
\begin{align}
    -M_{ii,+}^2&= -\frac{\mex_{t,ii}^2}{4},\quad-M_{ii,-}^2= -\infty\ ,& \text{ for }t\text{-channel case}\ ,\label{equ:-Mtii^2}\\
-M_{ii,+}^2&=-\frac{\mu_{+,i}^2}{4},\quad-M_{ii,-}^2=-\frac{\mu_{+,i}^2}{4\eta_i^2} \ , & \text{ for }u\text{-channel case}\ .\label{equ:-Muii^2}
\end{align}
Thus, Eq.~\eqref{equ:ImLij} naturally incorporates the single-channel results given in Eqs.~\eqref{equ:ImLtii} and~\eqref{equ:ImLuii}.

It should be noted that, according to the analysis in Appendix~\ref{sec:appendixofsingularity}, where the exchanged particle mass is so small that it nearly reaches the point where three-body kinematics would emerge at the two-body threshold, the lhc originates precisely at the lowest threshold value (Cases 2--4 in Fig.~\ref{fig:Singularityfortwochannel_all}), indicating that a massive system can develop an lhc structure analogous to that of a massless particle (e.g., a photon). These cases are rather special and beyond the present analysis.

\subsection{Singularity structure of the unitarized coupled-channel amplitude with the left-hand cut}
We now examine the singularity structure of the $\nch$-channel $S$-wave unitarized amplitude incorporating the OPE lhc from Eq.~\eqref{equLt_nonreal}, where $\nch$ is the number of coupling channels. From unitarity, the $S$-wave coupled-channel scattering amplitude $\mathbb{T}$ exhibits right-hand cut (rhc), which is characterized by
\begin{align}
    {\rm Im}\ \mathbb{T}_{\alpha\beta}^{-1}(k_\beta^2)=\mathbb{K}_{\alpha\beta}(k_\beta^2)\ ,\qquad \text{for }k_\beta^2>0\ ,\label{equ:unitaritystructure}
\end{align}
where
\begin{align}
    \mathbb{K}=\frac{1}{2\pi}\begin{pmatrix}
\mu_1k_1 & 0 & \cdots & 0 \\
0 & \mu_2k_2 & \cdots & 0 \\
\vdots & \vdots & \ddots & \vdots \\
0 & 0 & \cdots & \mu_\nch k_{\nch}
\end{pmatrix}.\label{equ:K}
\end{align}
The derivation in Refs.~\cite{Du:2024gzw,Wang:2026ups} uses that the discontinuity of $\mathbb{T}$ across the OPE lhc equals that of the OPE potential given in Eq.~\eqref{equLt_nonreal} when the lhcs due to the iterations are neglected, which is a very good approximation (see Fig.~\ref{fig:T[00]} below).
That is, 
\begin{align}
    {\rm Im}\ \mathbb{T}_{\alpha\beta}(k_\beta^2)={\rm Im}\ \mathbb{L}_{\alpha\beta}(k_\beta^2)\ , \qquad \text{for }k_\beta^2\in\left[ -M_{\beta\alpha,-}^2,-M_{\beta\alpha,+}^2 \right]\ ,\label{equ:lhcstructure}
\end{align}
where
\begin{align}
\mathbb{L}=\begin{pmatrix}
c_{11}L_{11}(k_1^2) & c_{12}L_{12}(k_2^2) & \cdots & c_{1{\nch}}L_{1{\nch}}(k_{\nch}^2) \\
c_{21}L_{21}(k_1^2) & c_{22}L_{22}(k_2^2) & \cdots & c_{2{\nch}}L_{2{\nch}}(k_{\nch}^2) \\
\vdots & \vdots & \ddots & \vdots \\
c_{{\nch}1}L_{{\nch}1}(k_1^2) & c_{{\nch}2}L_{{\nch}2}(k_2^2) & \cdots & c_{{\nch}{\nch}}L_{{\nch}{\nch}}(k_{\nch}^2) \label{equ:formofL}
\end{pmatrix}
\end{align}
with $c_{\alpha\beta}$ denoting the OPE strengths related to the product of coupling constants associated with the exchanged particle and the scattering particles.\footnote{In this work, we focus on the $S$-wave couplings between the scattering and the exchanged particles. Couplings in higher partial waves do not affect our analysis or results, as demonstrated in Ref.~\cite{Du:2024gzw}.} Time-reversal invariance imposes $c_{\alpha\beta}=c_{\beta\alpha}$ along with $L_{\alpha\beta}(k_\beta^2)=L_{\beta\alpha}(k_\alpha^2)$. 

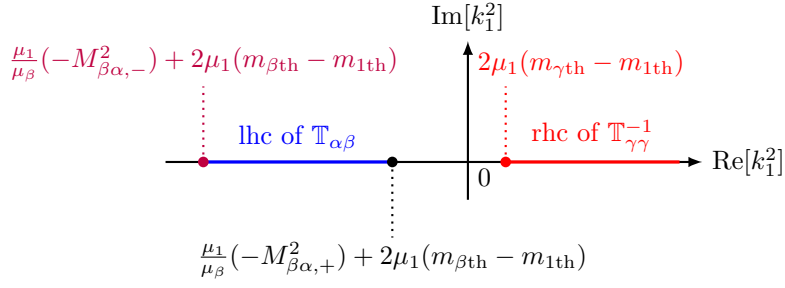
\begin{figure}[tb]
\centering
\begin{tikzpicture}[scale=1.0]
\begin{scope}
\draw[-latex,thick] (-4,0) -- (0,0)node [below=6pt, right,black] {0} -- (3.1,0) node[right] {${\rm Re}[k_1^2]$};
\draw[-latex,thick] (0,-0.5) --  (0,1.6) node[above] {${\rm Im}[k_1^2]$};
\draw[blue,very thick] (-3.5,0) --node[above]{lhc of $\mathbb{T}_{\alpha\beta}$} (-1,0) ;
\fill[fill=purple] (-3.5,0) circle[radius=2pt];
\draw[purple,thick,dotted] (-3.5,0) -- (-3.5,1);
\node[purple] at (-3.5,1.3 ) {$\frac{\mu_1}{\mu_\beta}(-M_{\beta\alpha,-}^2)+2\mu_1(m_{\beta{\rm th}}-m_{1{\rm th}})$};

\draw[black,thick,dotted] (-1,0) -- (-1,-1);
\fill[fill=black] (-1,0) circle[radius=2pt];
\node[black] at (-1,-1.3) {$\frac{\mu_1}{\mu_\beta}(-M_{\beta\alpha,+}^2)+2\mu_1(m_{\beta{\rm th}}-m_{1{\rm th}})$};

\fill[fill=red] (0.5,0) circle[radius=2pt];
\draw[red,very thick] (0.5,0) --node[above]{rhc of $\mathbb{T}^{-1}_{\gamma\gamma}$} (2.8,0);
\draw[red,thick,dotted] (0.5,0) -- (0.5,1);
\node[red] at (1.5,1.3) { $2\mu_1(m_{\gamma{\rm th}}-m_{1{\rm th}})$ };
\end{scope}
\end{tikzpicture}
\caption{Singularity structure of each matrix element of the ${\nch}$-channel $S$-wave unitarized amplitude $\mathbb{T}$ with the OPE lhc in the complex $k_1^2$-plane, based on the rhc and lhc defined in Eqs.~\eqref{equ:unitaritystructure} and~\eqref{equ:lhcstructure}, respectively.}
\label{fig:SingularityforF}
\end{figure}

From Eqs.~\eqref{equ:unitaritystructure} and~\eqref{equ:lhcstructure}, the singularity structure of $\mathbb{T}$ in the complex $k_1^2$-plane is depicted in Fig.~\ref{fig:SingularityforF}. According to Eq.~\eqref{relationkikj}, the complex \(k_i^2\) planes for other channels are simply globally shifted and scaled copies of the $k_1^2$-plane. In general, the lhc may overlap the rhc. It corresponds to the black dot lying to the right of the red dot in~Fig.~\ref{fig:SingularityforF}. In such a case, Eqs.~\eqref{equ:unitaritystructure} and~\eqref{equ:lhcstructure} would be invalid in the overlapping region and would hold only where the cuts are separated. 
To avoid this complication, in this work we focus on the case
\begin{align}
    {\rm max}\left\{\frac{-M_{\beta\alpha,+}^2}{2\mu_\beta}+m_{\beta{\rm th}}-m_{1{\rm th}}\right\}\leq0\ , \qquad \text{for }\alpha,\beta=1,2,\dots,{\nch}\ ,\label{equ:nooverlap}
\end{align}
which, together with Eq.~\eqref{equ:condition}, serves as the basis for our subsequent dispersion analysis. 

\begin{figure}[tb]
    \centering
    \includegraphics[width=0.68\textwidth]{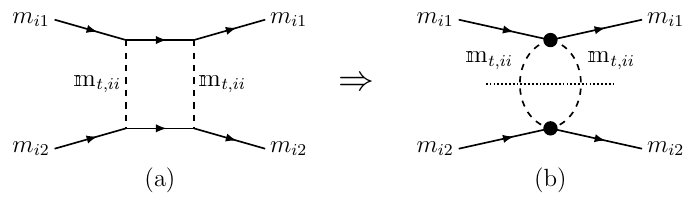}
    \caption{(a) Once-iterated contribution to the LSE for channel-$i$ in the single-channel situation. (b) Shrinking the two $s$-channel propagators reduces the diagram to a bubble diagram. The $t$-channel threshold gives rise to the new lhc branch point at $k_i^2 = -m_{t,ii}^2$.}
    \label{fig:once-iterated}
\end{figure}

Here we make a remark on new lhcs due to iterations of the OPE potential.
The branch points from iterations are further away from the threshold. 
For diagonal $t$-channel exchange, the once-iterated contribution with channel-$i$ in the loop starts at $k_i^2=-\mex_{t,ii}^2$, four times farther from threshold than the tree-level branch point. It is a subsubleading Landau singularity~\cite{Landau:1959fi} of the box diagram; see Fig.~\ref{fig:once-iterated}. 
For $k_i=i\kappa_i$ with $\kappa_i>\mex_{t,ii}$, the magnitude of its lhc discontinuity relative to the tree-level one can be worked out as 
\begin{align}
    \frac{\mu_i|c_{ii}|}{2\pi\kappa_i}\log\left(\frac{2\kappa_i}{\mex_{t,ii}}-1\right),
\end{align}
which agrees with the single-channel expression for nucleon-nucleon scattering in Ref.~\cite{Entem:2016ipb}.
It vanishes at this new branch point in addition to the $1/(2\pi)$ suppression factor. This supports retaining the nearest OPE lhc as a very good approximation in the near-threshold parameterization of Refs.~\cite{Du:2024gzw,Wang:2026ups}, as can also be seen from Fig.~\ref{fig:T[00]} below. 
Throughout the work, lhcs from these iterations are neglected.

\subsection{Dispersion construction of the parameterization for ${\nch}$-coupled channels}\label{sec:N/Dformula}

To simultaneously incorporate the OPE-induced lhc and the unitarity cut, we follow Refs.~\cite{Du:2024gzw,Du:2025vkm} and employ the $N/D$ method~\cite{Chew:1960iv,Bjorken:1960zz,Oller:2018zts,Oller:2019rej}. In this approach, the $S$-wave coupled-channel scattering amplitude can be expressed as
\begin{equation}
\mathbb{T} = \mathbb{D}^{-1}\,\mathbb{N}\ ,\label{eq:f=n/d}
\end{equation}
where $\mathbb{D}$ and $\mathbb{N}$ are ${\nch}\times {\nch}$ matrices. The matrix $\mathbb{D}$ contains only the rhc arising from unitarity, while $\mathbb{N}$ contains only the lhc. 
When only the OPE lhc is concerned, with Eqs.~\eqref{equ:unitaritystructure} and~\eqref{equ:lhcstructure}, the discontinuity relations across these cuts are given by
\iffalse
\begin{align}
    \text{Im }\mathbb{D}_{\alpha\beta}(k_\beta^2)&=\mathbb{N}_{\alpha\beta}(k_\beta^2)\,\mathbb{K}_{\beta\beta}(k_\beta^2), & \text{for }& k_\beta^2>0\ ,\label{ImofD}\\
    {\text{Im }}\mathbb{N}_{\alpha\beta}(k_\beta^2)&=\mathbb{D}_{\alpha\gamma}(k_\beta^2)\,{\text{Im }}\mathbb{L}_{\gamma\beta}(k_\beta^2), & \text{for }& \begin{cases}
   k_\beta^2\in[-M_{\beta\gamma,2}^2,-M_{\beta\gamma,1}^2]\ ,& \text{ if }\gamma>\beta \\
   k_\beta^2\in[-M_{\gamma\beta,2}^2,-M_{\gamma\beta,1}^2]\ ,& \text{ if }\gamma<\beta
   \end{cases}, \label{ImofN}
\end{align}
\fi
\begin{align}
    \text{Im }\mathbb{D}_{\alpha\beta}(k_\beta^2)&=\mathbb{N}_{\alpha\beta}(k_\beta^2)\,\mathbb{K}_{\beta\beta}(k_\beta^2)\ , & \text{for }& k_\beta^2>0\ ,\label{ImofD}\\
    {\text{Im }}\mathbb{N}_{\alpha\beta}(k_\beta^2)&=\sum_{\gamma=1}^{{\nch}}\mathbb{D}_{\alpha\gamma}(k_\gamma^2)\,{\text{Im }}\mathbb{L}_{\gamma\beta}(k_\beta^2)\ , & \text{for }& k_\beta^2 < 2\mu_\beta(m_{1{\rm th}}-m_{\beta{\rm th}}). \label{ImofN}
\end{align}
Applying dispersion relations to Eqs.~(\ref{ImofD}, \ref{ImofN}) yields the following integral representations~\cite{Du:2024gzw}:
\begin{align}\label{eq:nd_disp}
    \mathbb{D}_{\alpha\beta}(k_\beta^2)&=\tilde{\mathbb{D}}_{\alpha\beta,n}(k_\beta^2) + \frac{(k_\beta^2-k_{\beta,0}^2)^n}{\pi}\int_0^{\infty}\frac{\mathbb{N}_{\alpha\beta}(k_\beta'^2)\mathbb{K}_{\beta\beta}(k_\beta'^2){\rm d}k_\beta'^2}{(k_\beta'^2-k_\beta^2)(k_\beta'^2-k_{\beta,0}^2)^n}\ ,\\
    \mathbb{N}_{\alpha\beta}(k_\beta^2)&=\tilde{\mathbb{N}}_{\alpha\beta,m}(k_\beta^2) + \frac{(k_\beta^2)^m}{\pi}\sum_{\gamma=1}^{{\nch}}\int_{-M_{\beta\gamma,-}^2}^{-M_{\beta\gamma,+}^2}\frac{\mathbb{D}_{\alpha\gamma}(k_\gamma'^2){\rm Im}\,\mathbb{L}_{\gamma\beta}(k_\beta'^2)}{(k_\beta'^2-k_\beta^2)(k_\beta'^2)^m}{\rm d}k_\beta'^2\ .
\end{align}
where $m$, $n$ are the numbers of subtractions required to render the integrals finite, $k_{\beta,0}^2$ is the subtraction point, and $\tilde{\mathbb{D}}_{\alpha\beta,n}(k_\beta^2)$ and $\tilde{\mathbb{N}}_{\alpha\beta,m}(k_\beta^2)$ are polynomials of $k_\beta^2$. Using the method developed in Refs.~\cite{Du:2024gzw,Wang:2026ups} and the singularity structure of the $S$-wave OPE potential discussed in Sec.~\ref{singularityofL}, and noting that Eq.~\eqref{eq:f=n/d} implies that $\mathbb{N}$ and $\mathbb{D}$ can be simultaneously multiplied on the left by any invertible matrix $\mathbb{X}$ without changing $\mathbb{T}$, i.e., 
\begin{align}
    \mathbb{T}=(\mathbb{X}\mathbb{D})^{-1}(\mathbb{X}\mathbb{N})\ ,\label{equ:F=XDinv_XN}
\end{align}
we can evaluate the dispersion integrals to obtain a form similar to the single-channel case in Refs.~\cite{Du:2024gzw,Wang:2026ups}. Specifically, we write
\begin{align}
    \mathbb{N}_{\alpha\beta}(k_\beta^2)&=\tilde{\mathbb{N}}_{\alpha\beta}(k_\beta^2) + \sum_{\gamma=1}^{{\nch}}\tilde{\mathbb{G}}_{\alpha\gamma}\mathbb{L}_{\gamma\beta}(k_\beta^2)\ ,\label{resultofN}\\
    \mathbb{D}_{\alpha\beta}(k_\beta^2)&=\tilde{\mathbb{D}}_{\alpha\beta}(k_\beta^2)+i\,\mathbb{N}_{\alpha\beta}(k_\beta^2)\mathbb{K}_{\beta\beta}(k_\beta^2) + \sum_{\gamma=1}^{{\nch}}\tilde{\mathbb{G}}_{\alpha\gamma}\mathbb{D}_{\gamma\beta}^{\rm R}(k_\beta^2)\ ,\label{resultofD}
\end{align}
with
\begin{align}
    \mathbb{D}_{\gamma\beta}^{\rm R}(k_\beta^2)=-\frac{i}{\pi}\int_{-M_{\beta\gamma,-}^2}^{-M_{\beta\gamma,+}^2}\frac{\mathbb{K}_{\beta\beta}(k_\beta'^2)\,\text{Im }\mathbb{L}_{\gamma\beta}(k_\beta'^2)}{k_\beta'^2-k_\beta^2}{\rm d}k_\beta'^2\ ,\label{equ:DRdefine}
\end{align}
and $\tilde{\mathbb{N}}_{\alpha\beta}(k_\beta^2)$ and $\tilde{\mathbb{D}}_{\alpha\beta}(k_\beta^2)$ are, in general, rational functions of $k_\beta^2$.
The term $\mathbb{D}^{\rm R}$ cancels the lhc introduced by $\mathbb{N}$ in Eq.~(\ref{resultofD}), thereby ensuring that $\mathbb{D}$ contain only rhc, namely Eq.~(\ref{ImofD}). Using the formulas in Appendix~\ref{sec:calofdR}, explicit expressions for $\mathbb{D}^{\rm R}$ can be derived as
\begin{align}
    \mathbb{D}_{\gamma\beta}^{\rm R}(k_\beta^2)= 
   \frac{ic_{\gamma\beta}\mu_\beta}{8\pi k_\gamma}\log\left(\frac{ik_\gamma-i\sqrt{-M_{\gamma\beta,+}^2}}{ik_\gamma+i\sqrt{-M_{\gamma\beta,+}^2}}\frac{ik_\gamma+i\sqrt{-M_{\gamma\beta,-}^2}}{ik_\gamma-i\sqrt{-M_{\gamma\beta,-}^2}} \right).\label{equ:DRresult}
\end{align}

One can make use of the freedom in the normalization of the numerator and denominator in the quotient by choosing $\mathbb{X} = \tilde{\mathbb{G}}^{-1}$ in Eq.~\eqref{equ:F=XDinv_XN}. With this choice, Eqs.~\eqref{resultofN} and \eqref{resultofD} simplify to
\begin{align}
\mathbb{N}_{\alpha\beta}(k_\beta^2)&=\tilde{\mathbb{N}}_{\alpha\beta}(k_\beta^2) + \mathbb{L}_{\alpha\beta}(k_\beta^2)\ ,\label{resultofNwithoutG}\\
\mathbb{D}_{\alpha\beta}(k_\beta^2)&=\tilde{\mathbb{D}}_{\alpha\beta}(k_\beta^2)+i\,\mathbb{N}_{\alpha\beta}(k_\beta^2)\mathbb{K}_{\beta\beta}(k_\beta^2) + \mathbb{D}_{\alpha\beta}^{\rm R}(k_\beta^2)\ .\label{resultofDwithout}
\end{align}
In practice, a more efficient parametrization of the amplitude is obtained by expanding these rational functions $\tilde{\mathbb{N}}$ and $\tilde{\mathbb{D}}$ in Taylor series and truncating them at orders $m$ and $n$~\cite {Du:2024gzw,Du:2025vkm,Wang:2026ups}, respectively.
For that, we first define
\begin{align}
\mathbb{Q}_{[m,n]}=\left(\tilde{\mathbb{N}}_m+\mathbb{L}\right)^{-1}\left( \tilde{\mathbb{D}}_{n}+\mathbb{D}^{\rm R}\right) \ ,\label{equ:Q}
\end{align}
where $\tilde{\mathbb{N}}_{m}$ and $\tilde{\mathbb{D}}_{n}$ are matrices constructed from Taylor power series in $k_1^2$.\footnote{Owing to the linear relation in Eq.~(\ref{relationkikj}), the square of the three-momentum of any channel can be chosen as the expansion variable; this choice has virtually no effect on the final result.} Explicitly,
\begin{align}
    \tilde{\mathbb{N}}_{m,\alpha\beta}(k_\beta^2)=\sum_{j=0}^{m}\tilde{\mathbb{N}}_{j,\alpha\beta}k_1^{2j}\ , \quad \tilde{\mathbb{D}}_{n,\alpha\beta}(k_\beta^2)&=\sum_{j=0}^{n}\tilde{\mathbb{D}}_{j,\alpha\beta}k_1^{2j}\ .
\end{align}

Two remarks are in order. First, comparing Eqs.~(\ref{ImofN}) and (\ref{resultofNwithoutG}) reveals that this normalization scheme ideally requires $\mathbb{D}$ to be an identity matrix along the OPE lhc region. However, the full $\mathbb{D}$ contains the energy-dependent dispersive terms in Eq.~\eqref{resultofDwithout}, which cannot in general be canceled by a finite polynomial $\tilde{\mathbb{D}}$. We therefore impose the normalization only at the lhc branch point:
\begin{align}
    \mathbb{D}_{\alpha\beta}(k_\beta^2=-M_{\beta\alpha,+}^2)=\delta_{\alpha\beta}\ .\label{equ:Disunit}
\end{align}
Second, time-reversal invariance requires the complete on-shell amplitude $\mathbb{T}$ to be symmetric. We take the polynomial matrices $\tilde{\mathbb{N}}_m$ and $\tilde{\mathbb{D}}_n$ with $n\ge1$ to be real symmetric:
\begin{align}
  \tilde{\mathbb{N}}_m^T=\tilde{\mathbb{N}}_m\ ,
  \qquad 
  \tilde{\mathbb{D}}_n^T=\tilde{\mathbb{D}}_n \quad (n\ge1)
  \label{equ:Nmsymmetric}
\end{align}
and also symmetrize the matrix $\mathbb{Q}_{[m,n]}$ in Eq.~\eqref{equ:Q}\footnote{Note that $\mathbb{Q}_{[m,n]}=\left(\tilde{\mathbb{N}}_m+\mathbb{L}\right)^{-1}\left( \tilde{\mathbb{D}}_{n}+\mathbb{D}^{\rm R}\right)$ is not symmetric even if the two factors it contains are symmetric. It becomes symmetric only when these two factors commute.} to express the inverse amplitude as
\begin{align}
\mathbb{T}^{-1}_{[m,n]}=\frac12 \left(\mathbb{Q}_{[m,n]}+\mathbb{Q}_{[m,n]}^T\right)+i\,\mathbb{K}\ ,\label{equ:finalF}
\end{align}
which, being symmetric, satisfies time-reversal invariance.

The $[m,n]$ parametrization in Eq.~\eqref{equ:finalF} for an $N$-channel system contains 
\begin{align}
\frac{\nch^2+\nch}{2}\times(n+m+2)
\end{align}
free parameters. Compared with the conventional coupled-channel ERE in Eq.~\eqref{equ:CCERE_traditional}, our new parameterization merely adds $(\nch^2 + {\nch})/2$ OPE strength parameters $c_{\alpha\beta}$ which are related to the products of couplings between the exchanged and scattering particles.

Unlike in Refs.~\cite{Du:2024gzw,Liu:2026xrk}, we have normalized $\mathbb{D}$ as shown in Eq.~\eqref{equ:Disunit}, so that the product of coupling constants $c_{\alpha\beta}$ serve directly as fitting parameters. Thus, there is no need to be divided by the value of $\mathbb{D}$ at the lhc branch point, as was required in those references. In other words, the $\tilde{n}$ function is normalized to unity in Refs.~\cite{Du:2024gzw,Liu:2026xrk}, while the normalization of $\mathbb{D}$ is imposed at the lhc branch points according to Eq.~\eqref{equ:Disunit} in the present work.

\subsection{Explicit parameterization for two coupled channels}\label{sec:two_channel_case}
As an illustrative example, we now present the explicit parametrization for a two-channel case. For $S$-wave two-channel scattering, we assume that each matrix element receives a contribution from the $t$-channel OPE potential, i.e.,
\newcommand{\feynmanA}{
  \begin{tikzpicture}[baseline=(current bounding box.center), scale=0.8, every node/.style={scale=0.8}]
    \draw[ thick] (0,0) node[left] {$m_{11}$} -- (1,0);
    \draw[ thick] (1,0) -- (2,0) node[right] {$m_{11}$};
    \draw[ thick] (0,-1.25) node[left] {$m_{12}$} -- (1,-1.25);
    \draw[ thick] (1,-1.25) -- (2,-1.25) node[right] {$m_{12}$};
    \draw[dashed,thick] (1,0) -- node[left] {$\mex_{t,11}$} (1,-1.25);
  \end{tikzpicture}
}
\newcommand{\feynmanB}{
  \begin{tikzpicture}[baseline=(current bounding box.center), scale=0.8, every node/.style={scale=0.8}]
    \draw[ thick] (0,0) node[left] {$m_{11}$} -- (1,0);
    \draw[ thick] (1,0) -- (2,0) node[right] {$m_{21}$};
    \draw[ thick] (0,-1.25) node[left] {$m_{12}$} -- (1,-1.25);
    \draw[ thick] (1,-1.25) -- (2,-1.25) node[right] {$m_{22}$};
    \draw[dashed,thick] (1,0) -- node[left] {$\mex_{t,21}$} (1,-1.25);
  \end{tikzpicture}
}

\newcommand{\feynmanC}{
  \begin{tikzpicture}[baseline=(current bounding box.center), scale=0.8, every node/.style={scale=0.8}]
    \draw[ thick] (0,0) node[left] {$m_{21}$} -- (1,0);
    \draw[ thick] (1,0) -- (2,0) node[right] {$m_{11}$};
    \draw[ thick] (0,-1.25) node[left] {$m_{22}$} -- (1,-1.25);
    \draw[ thick] (1,-1.25) -- (2,-1.25) node[right] {$m_{12}$};
    \draw[dashed,thick] (1,0) -- node[left] {$\mex_{t,21}$} (1,-1.25);
  \end{tikzpicture}
}

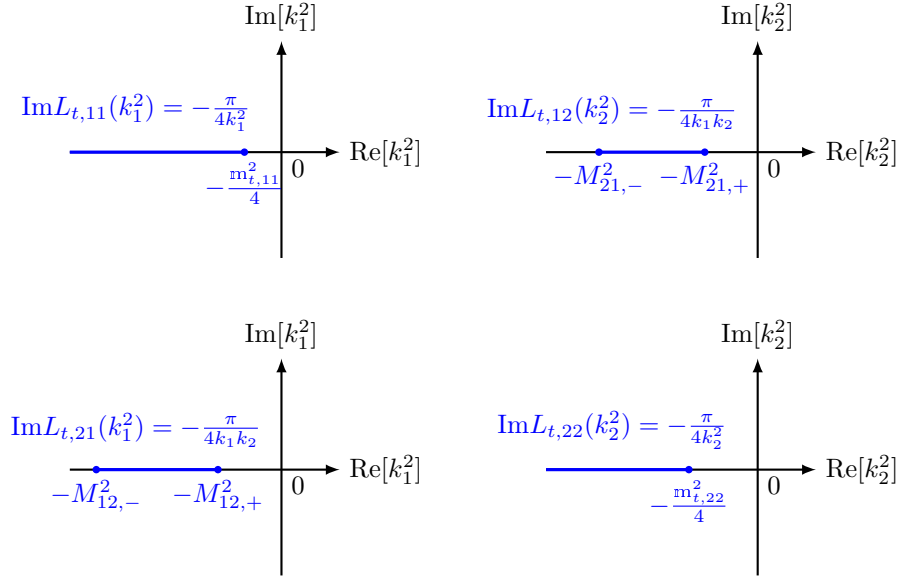
\begin{figure}[tb]
\centering
\begin{tikzpicture}[scale=0.7]
\begin{scope}
\draw[-latex,thick] (-4,0) -- (0,0) node [below=6pt, right,black] {0}-- (1.1,0) node[right] {${\rm Re}[k_1^2]$};
\draw[-latex,thick] (0,-2) --  (0,2.1) node[above] {${\rm Im}[k_1^2]$};
\draw[blue,very thick] (-4,0) -- (-0.7,0) node[below] {$-\frac{\mex_{t,11}^2}{4}$};
\fill[fill=blue] (-0.7,0) circle[radius=2pt];
\node[blue] at (-2.75,0.75) {Im$L_{t,11}(k_1^2)=-\frac{\pi}{4k_1^2}$};
\end{scope}

\begin{scope}[xshift=9cm]
\draw[-latex,thick] (-4,0) -- (0,0) node [below=6pt, right,black] {0}-- (1.1,0) node[right] {${\rm Re}[k_2^2]$};
\draw[-latex,thick] (0,-2) -- (0,2.1) node[above] {${\rm Im}[k_2^2]$};
\draw[blue,very thick] (-3,0) node[below] {$-M_{21,-}^2$} -- (-1,0) node[below] {$-M_{21,+}^2$};
\fill[fill=blue] (-1,0) circle[radius=2pt];
\fill[fill=blue] (-3,0) circle[radius=2pt];
\node[blue] at (-2.75,0.75) {Im$L_{t,12}(k_2^2)=-\frac{\pi}{4k_1k_2}$};
\end{scope}

\begin{scope}[yshift=-6cm]
\draw[-latex,thick] (-4,0) -- (0,0) node [below=6pt, right,black] {0}-- (1.1,0) node[right] {${\rm Re}[k_1^2]$};
\draw[-latex,thick] (0,-2) -- (0,2.1) node[above] {${\rm Im}[k_1^2]$};
\draw[blue,very thick] (-3.5,0) node[below] {$-M_{12,-}^2$} -- (-1.2,0) node[below] {$-M_{12,+}^2$};
\fill[fill=blue] (-1.2,0) circle[radius=2pt];
\fill[fill=blue] (-3.5,0) circle[radius=2pt];
\node[blue] at (-2.75,0.75) {Im$L_{t,21}(k_1^2)=-\frac{\pi}{4k_1k_2}$};
\end{scope}

\begin{scope}[xshift=9cm,yshift=-6cm]
\draw[-latex,thick] (-4,0) -- (0,0) node [below=6pt, right,black] {0}-- (1.1,0) node[right] {${\rm Re}[k_2^2]$};
\draw[-latex,thick] (0,-2) --  (0,2.1) node[above] {${\rm Im}[k_2^2]$};
\draw[blue,very thick] (-4,0) -- (-1.3,0) node[below] {$-\frac{\mex_{t,22}^2}{4}$};
\fill[fill=blue] (-1.3,0) circle[radius=2pt];
\node[blue] at (-2.75,0.75) {Im$L_{t,22}(k_2^2)=-\frac{\pi}{4k_2^2}$};
\end{scope}
\end{tikzpicture}
\caption{Singularity structure of the two-channel $S$-wave OPE potential defined in Eq.~\eqref{equ:Lintwochannel} in the complex $k_l^2$-plane ($l=1,2$), based on the lhc defined in Eqs.~\eqref{equ:ImL1122},~\eqref{equ:ImL21}, and~\eqref{equ:ImL12}.}
    \label{fig:Singularityfortwochannel}
\end{figure}

\newcommand{\feynmanD}{
  \begin{tikzpicture}[baseline=(current bounding box.center), scale=0.8, every node/.style={scale=0.8}]
    \draw[thick] (0,0) node[left] {$m_{21}$} -- (1,0);
    \draw[ thick] (1,0) -- (2,0) node[right] {$m_{21}$};
    \draw[ thick] (0,-1.25) node[left] {$m_{22}$} -- (1,-1.25);
    \draw[ thick] (1,-1.25) -- (2,-1.25) node[right] {$m_{22}$};
    \draw[dashed,thick] (1,0) -- node[left] {$\mex_{t,22}$} (1,-1.25);
  \end{tikzpicture}
}

\begin{align}
&\mathbb{L}^{(2)} =
\left(\begin{array}{cc}
  c_{11}L_{t,11}(k_1^2) & c_{12}L_{t,12}(k_2^2) \\
  c_{21}L_{t,21}(k_1^2) & c_{22}L_{t,22}(k_2^2)
\end{array}\right)= 
\left(\begin{array}{cc}
  \feynmanA & \feynmanB \\
  \feynmanC & \feynmanD 
\end{array}\right),
\label{equ:Lintwochannel}
\end{align}
where the superscript $(2)$ indicates that this matrix corresponds to the two-channel case, and time-reversal invariance imposes $c_{12}=c_{21}$ along with $L_{t,12}(k_2^2)=L_{t,21}(k_1  ^2)$. If a certain matrix element does not contain an OPE-induced lhc, one simply sets the corresponding coupling $c_{\alpha\beta}=0$ in Eq.~(\ref{equ:Lintwochannel}). Focusing on the most common scenario, characterized by the conditions in Eq.~\eqref{equ:condition}, in the two-channel case is
\begin{align} 
    \mex_{t,12}^2 > (m_{11}-m_{21})^2 + 2(m_{2\rm th}-m_{1\rm th})\frac{m_{11}}{m_{21}}\mu_2\ ,\label{equ:commom_for_two}
\end{align} 
and using Eqs.~(\ref{equ:ImLij}) and (\ref{equ:-Mtii^2}), we obtain the following discontinuity:
\begin{align}
    \text{Im }L_{t,ii}(k_i^2)&=-\frac{\pi}{4k_i^2},& &\text{ for }k_i^2<-\frac{\mex_{t,ii}^2}{4}\quad(i=1,2)\ ,\label{equ:ImL1122}\\
    \text{Im }L_{t,12}(k_2^2)&=-\frac{\pi}{4k_1k_2},& &\text{ for }k_2^2\in[-M_{21,-}^2,-M_{21,+}^2]\ .\label{equ:ImL21}\\
    \text{Im }L_{t,21}(k_1^2)&=-\frac{\pi}{4k_1k_2},& &\text{ for }k_1^2\in[-M_{12,-}^2,-M_{12,+}^2]\ .\label{equ:ImL12}
\end{align}
The resulting singularity structure of each component of $\mathbb{L}^{(2)}$ in the complex momentum-squared plane is illustrated schematically in Fig.~\ref{fig:Singularityfortwochannel}.

Using Eq.~\eqref{equ:DRresult}, we obtain the following explicit expression for $\mathbb{D}^{\mathrm{R},(2)}$:
\begin{align}
    \mathbb{D}^{\rm R,(2)}=\left(\begin{array}{cc}
  -\frac{ic_{11}\mu_1}{8\pi k_1}\log\left( \frac{\mex_{t,11}/2-ik_1}{\mex_{t,11}/2+ik_1} \right)  & \frac{ic_{21}\mu_2}{8\pi k_1}\log\left( \frac{ik_1-i\sqrt{-M_{12,+}^2}}{ik_1+i\sqrt{-M_{12,+}^2}}\frac{ik_1+i\sqrt{-M_{12,-}^2}}{ik_1-i\sqrt{-M_{12,-}^2}}\right)  \\
  \frac{ic_{21}\mu_1}{8\pi k_2}\log\left( \frac{ik_2-i\sqrt{-M_{21,+}^2}}{ik_2+i\sqrt{-M_{21,+}^2}}\frac{ik_2+i\sqrt{-M_{21,-}^2}}{ik_2-i\sqrt{-M_{21,-}^2}}\right)  &  -\frac{ic_{22}\mu_2}{8\pi k_2}\log\left( \frac{\mex_{t,22}/2-ik_2}{\mex_{t,22}/2+ik_2} \right)
\end{array}\right).\label{equ:DR2}
\end{align}

Substituting these results into Eq.~(\ref{equ:finalF}) yields the parametrization of the two-channel $S$-wave scattering amplitude:
\begin{align}
\left(\mathbb{T}^{(2)}_{[m,n]}\right)^{-1}&=\frac{1}{2}\left[\mathbb{Q}^{(2)}_{[m,n]}+\left(\mathbb{Q}^{(2)}_{[m,n]}\right)^T\right]+i\,\mathbb{K}^{(2)}\ ,\label{equ:Ffortwochannel}\\
\text{with } \mathbb{Q}^{(2)}_{[m,n]}&=\left(\tilde{\mathbb{N}}^{(2)}_m+\mathbb{L}^{(2)}\right)^{-1}\left( \tilde{\mathbb{D}}_{n}^{(2)}+\mathbb{D}^{\rm R,(2)}\right).
\end{align}
where, 
using Eqs.~(\ref{equ:K}, \ref{equ:Disunit}, \ref{equ:Nmsymmetric}), we have
\begin{align}
    \mathbb{K}^{(2)}=\frac{1}{2\pi}\left(\begin{array}{cc}
    \mu_1 k_1 & 0\\
    0 & \mu_2 k_2
     \end{array}\right), \  \tilde{\mathbb{N}}_m^{(2)}=\sum_{j=0}^m\left(\begin{array}{cc}
     \tilde{\mathbb{N}}_{j,11} & \tilde{\mathbb{N}}_{j,12} \\
     \tilde{\mathbb{N}}_{j,12} & \tilde{\mathbb{N}}_{j,22} 
     \end{array}\right)k_1^{2j},\  \tilde{\mathbb{D}}^{(2)}_n=\tilde{\mathbb{D}}_0+\sum_{j=1}^{n}\left(\begin{array}{cc}
     \tilde{\mathbb{D}}_{j,11} & \tilde{\mathbb{D}}_{j,12} \\
     \tilde{\mathbb{D}}_{j,21} & \tilde{\mathbb{D}}_{j,22} 
     \end{array}\right)k_1^{2j}.
\end{align}
The constant matrix $\tilde{\mathbb{D}}_0$ is given explicitly by
\begin{equation}
\begin{aligned}
\tilde{\mathbb{D}}_0=&\left(\begin{array}{cc}
     1 & 0 \\
     0 & 1 
\end{array}\right)-\sum\limits_{j=1}^{n}\left(\begin{array}{cc}
     \tilde{\mathbb{D}}_{j,11}\left(-\frac{\mex_{t,11}^2}{4}\right)^j & \tilde{\mathbb{D}}_{j,12}\left(-M_{12,+}^2\right)^j \\
     \tilde{\mathbb{D}}_{j,21}\left(-M_{12,+}^2\right)^j & \tilde{\mathbb{D}}_{j,22}H_{22}^j 
     \end{array}\right)\\
&+\left(
\begin{array}{cc}
     \frac{c_{11}\mu_1}{2\pi\mex_{t,11}}\log 2 & \frac{-c_{21}\mu_2}{8\pi\sqrt{M_{12,+}^2}}\log\left[H_{12}\right] \\
     \frac{-c_{21}\mu_1}{8\pi\sqrt{M_{21,+}^2}}\log\left[\frac{\mu_2\sqrt{-M_{12,+}^2}H_{21}}{\mu_1\sqrt{-M_{21,+}^2}}\right] & \frac{c_{22}\mu_2}{2\pi\mex_{t,22}}\log 2
     \end{array}\right)\\
     &-i\sum\limits_{j=0}^{m}\left(\begin{array}{cc}
\frac{i\mu_1\mex_{t,11}}{4\pi}\tilde{\mathbb{N}}_{j,11}\left(-\frac{\mex_{t,11}^2}{4}\right)^j & \frac{\mu_2\sqrt{-M_{21,+}^2} }{2\pi}\tilde{\mathbb{N}}_{j,12}\left(-M_{12,+}^2\right)^j \\
\frac{\mu_1\sqrt{-M_{12,+}^2} }{2\pi}\tilde{\mathbb{N}}_{j,12}\left(-M_{12,+}^2\right)^j & \frac{i\mu_2\mex_{t,22}}{4\pi}\tilde{\mathbb{N}}_{j,22}H_{22}^j 
\end{array}\right),
\end{aligned}
\end{equation}
where we have introduced the shorthand notations
\begin{align}
H_{12}&=h(-M_{12,+}^2,-M_{21,+}^2)\, g(-M_{12,+}^2,-M_{12,-}^2)\ ,\notag\\
H_{21}&=h(-M_{12,+}^2,-M_{21,+}^2)\, g\left(-M_{21,+}^2,-M_{21,-}^2\right) ,\notag\\
H_{22}&=\frac{\mu_1}{\mu_2}\left(-\frac{\mex_{t,22}^2}{4}\right)+2\mu_1(m_{2{\rm th}}-m_{1{\rm th}})\ .
\notag
\end{align}
The auxiliary functions $g(x,y)$ and $h(x,y)$ are defined as
\begin{align}
    g(x,y)&=\frac{\sqrt{x}+\sqrt{y}}{4\sqrt{x}\left(\sqrt{x}-\sqrt{y}\right)}\ ,\quad h(x,y)=\frac{\left(\sqrt{\frac{m_{21}}{m_{11}}}\sqrt{x}-\sqrt{\frac{m_{11}}{m_{21}}}\sqrt{y} \right)^2+\mex_{t,21}^2-(m_{11}-m_{21})^2}{\left(\sqrt{\frac{m_{21}}{m_{11}}}\sqrt{x}+\sqrt{\frac{m_{11}}{m_{21}}}\sqrt{y}\right)\left(\sqrt{\frac{m_{21}}{m_{11}}}+\sqrt{\frac{m_{11}}{m_{21}}}\frac{\mu_2\sqrt{x}}{\mu_1\sqrt{y}}\right)}\ .\notag
\end{align}
Moreover, the $\tilde{\mathbb{D}}_j$ matrices are taken to be real symmetric as mentioned above. With this choice, the $[m,n]$ parametrization presented in Eq.~\eqref{equ:Ffortwochannel} contains a total of $3(n+m+2)$ free parameters. The lowest order $(\mathbb{T}^{(2)}_{[0,0]})^{-1}=\big(\mathbb{Q}^{(2)}_{[0,0]}+(\mathbb{Q}^{(2)}_{[0,0]})^T\big)/2+i\mathbb{K}^{(2)}$ contains only six parameters:
\begin{align}
    \mathbb{Q}^{(2)}_{[0,0]}= \left( \left(\begin{array}{cc}
     \tilde{\mathbb{N}}_{0,11} & \tilde{\mathbb{N}}_{0,12} \\
     \tilde{\mathbb{N}}_{0,12} & \tilde{\mathbb{N}}_{0,22}
     \end{array}\right)+\mathbb{L}^{(2)}\right)^{-1}\left( \left( \begin{array}{cc}
      \tilde{d}_{11} &\tilde{d}_{12}  \\
      \tilde{d}_{21} &\tilde{d}_{22} 
     \end{array}\right)+\mathbb{D}^{\rm R,(2)} \right) ,\label{equ:F00}
\end{align}
where the explicit forms of $\mathbb{L}^{(2)}$ and $\mathbb{D}^{\rm R,(2)}$ are given in Eqs.~\eqref{equ:Lintwochannel} and~\eqref{equ:DR2}, respectively, and
\begin{align}
    \tilde{d}_{11}&=1+\frac{\mu_1\mex_{t,11}}{4\pi}\tilde{\mathbb{N}}_{0,11}+\frac{c_{11}\mu_1}{2\pi\mex_{t,11}}\log2\ ,\\
    \tilde{d}_{12}&=-i\frac{\mu_2\sqrt{-M_{21,+}^2}}{2\pi}\tilde{\mathbb{N}}_{0,12}-\frac{c_{21}\mu_2}{8\pi\sqrt{M_{12,+}^2}}\log\left[H_{12}\right]\ ,\\
    \tilde{d}_{21}&=-i\frac{\mu_1\sqrt{-M_{12,+}^2}}{2\pi}\tilde{\mathbb{N}}_{0,12}-\frac{c_{21}\mu_1}{8\pi\sqrt{M_{21,+}^2}}\log\left[\frac{\mu_2\sqrt{-M_{12,+}^2}H_{21}}{\mu_1\sqrt{-M_{21,+}^2}}\right]\ ,\\
    \tilde{d}_{22}&=1+\frac{\mu_2\mex_{t,22}}{4\pi}\tilde{\mathbb{N}}_{0,22}+\frac{c_{22}\mu_2}{2\pi\mex_{t,22}}\log 2\ .
\end{align}
When the OPE lhc is absent, i.e., $c_{11} =c_{21}=c_{22}= 0$, the $[0,0]$ approximant in Eq.~\eqref{equ:F00} reduces to the conventional coupled-channel ERE in Eq.~\eqref{equ:CCERE_traditional}.

\section{Results and discussion}\label{sec:results}

In this section, we demonstrate the efficacy of the proposed coupled-channel parametrization of the scattering amplitude. To this end, we compare the parametrization given in Eq.~(\ref{equ:Ffortwochannel}) with the scattering amplitudes obtained by numerically solving the coupled-channel LSE that incorporates the OPE potentials.

\subsection{Coupled-channel amplitude from LSE}
Starting from the off-shell potential in Eq.~(\ref{equ:Lintwochannel}), the two-channel $S$-wave scattering amplitude $\mathbb{T}$ can be obtained by solving the coupled-channel nonrelativistic LSE (see, e.g., Ref.~\cite{Du:2021zzh})
\begin{align}\label{equ:LSE}
\mathbb{T}_{ij}(E,p_i,q_j)=\mathbb{V}_{ij}(p_i,q_j)+\sum_{\alpha=1}^{2}\int_0^{\infty} \frac{{\rm d}l_\alpha}{2\pi^2} \frac{l_\alpha^2\mathbb{V}_{i\alpha}(p_i,l_\alpha)\mathbb{T}_{\alpha j}(E,l_\alpha,q_j)}{E-l_\alpha^{\,2}/2\mu_\alpha-(m_{\alpha\rm th}-m_{1\rm th}) + i\epsilon}\ ,
\end{align}
where the off-shell potential is given by
\begin{align}
    \mathbb{V}_{ij}(p_i,q_j)=\frac{c_{ij}}{4p_iq_j}\log\left[ \frac{\mex_{t,ij}^2-(m_{i1}-m_{j1})^2+\left( \sqrt{\frac{m_{j1}}{m_{i1}}}p_i-\sqrt{\frac{m_{i1}}{m_{j1}}}q_j\right)^2}{\mex_{t,ij}^2-(m_{i1}-m_{j1})^2+\left( \sqrt{\frac{m_{j1}}{m_{i1}}}p_i+\sqrt{\frac{m_{i1}}{m_{j1}}}q_j\right)^2} \right].\label{equ:VinLSE}
\end{align}
Here, $E$ denotes the energy measured relative to the threshold of the first channel, while $p_i$ and $q_j$ represent the magnitudes of the three-momenta in the respective channels. 
It should be noted that the integral term in the LSE~\eqref{equ:LSE} is ultraviolet convergent as the potential $\mathbb{V}_{i\alpha}(p_i,l_\alpha)$ in Eq.~\eqref{equ:VinLSE} scales as $l_\alpha^{-2}$ for large $l_\alpha$. Hence, no regulator is required.

\begin{figure}[tb]
    \centering
    \begin{minipage}{0.499\linewidth}
        \centering
        \begin{overpic}[width=\linewidth]{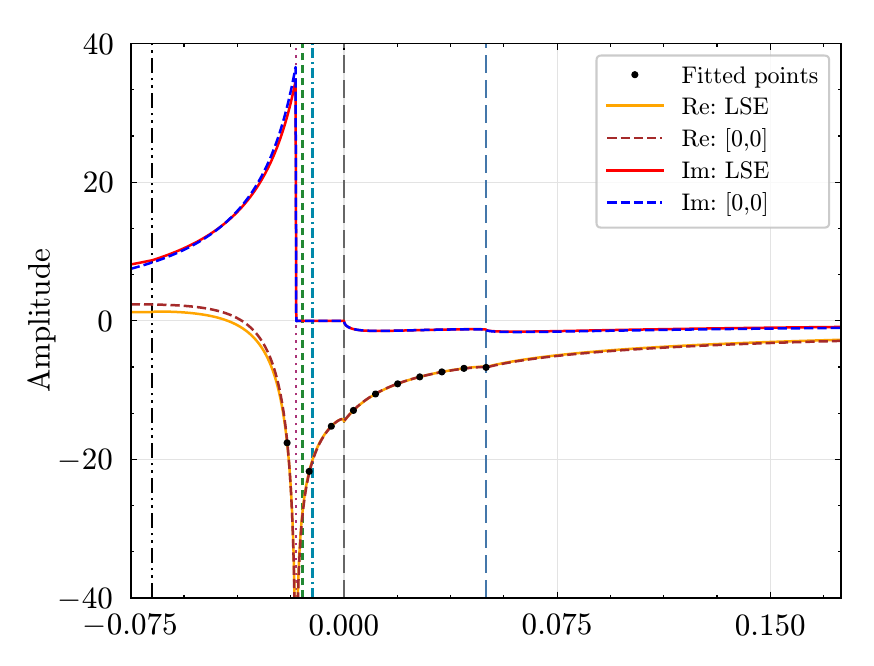} 
            \put(70,20){$\mathbb{T}_{11}$}  
        \end{overpic}
    \end{minipage}
    \hfill
    \begin{minipage}{0.479\linewidth}
        \centering
        \begin{overpic}[width=\linewidth]{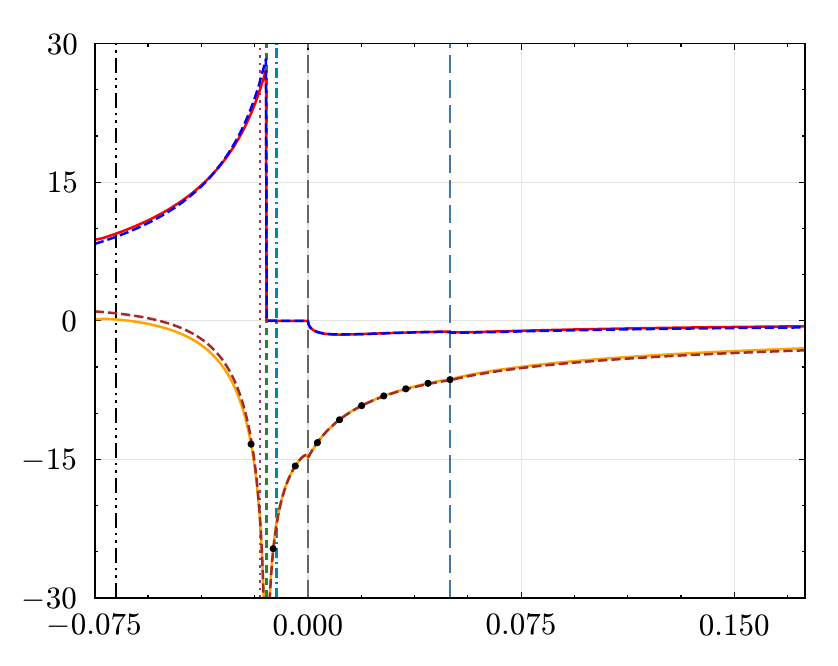}
            \put(70,20){$\mathbb{T}_{12}$}
        \end{overpic}
    \end{minipage}
    \\[3mm]
    \begin{minipage}{0.512\linewidth}
        \centering
        \begin{overpic}[width=\linewidth]{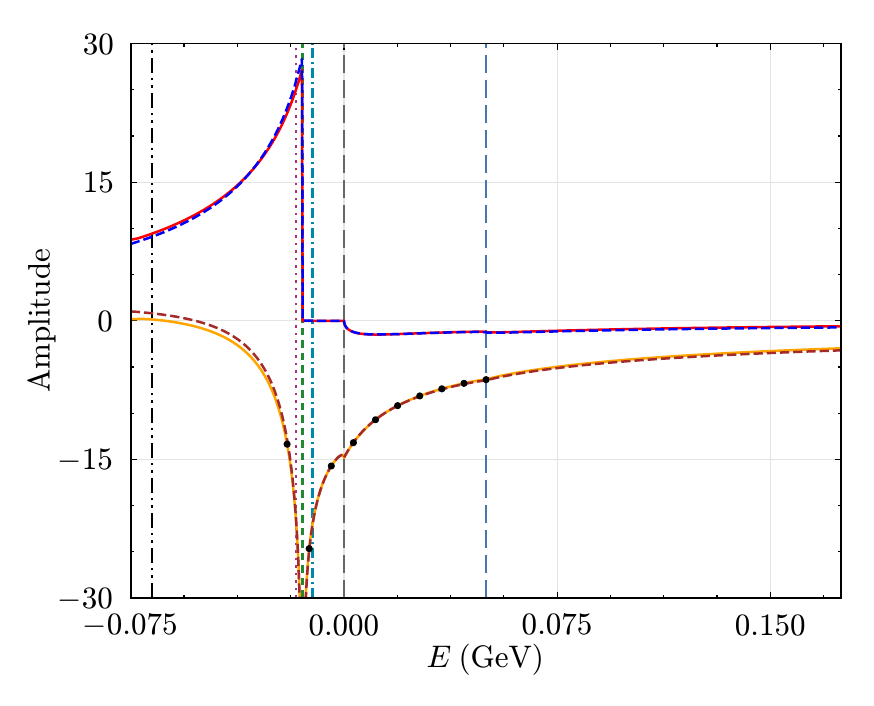}
            \put(70,25){$\mathbb{T}_{21}$}
        \end{overpic}
    \end{minipage}
    \hfill
    \begin{minipage}{0.469\linewidth}
        \centering
        \begin{overpic}[width=\linewidth]{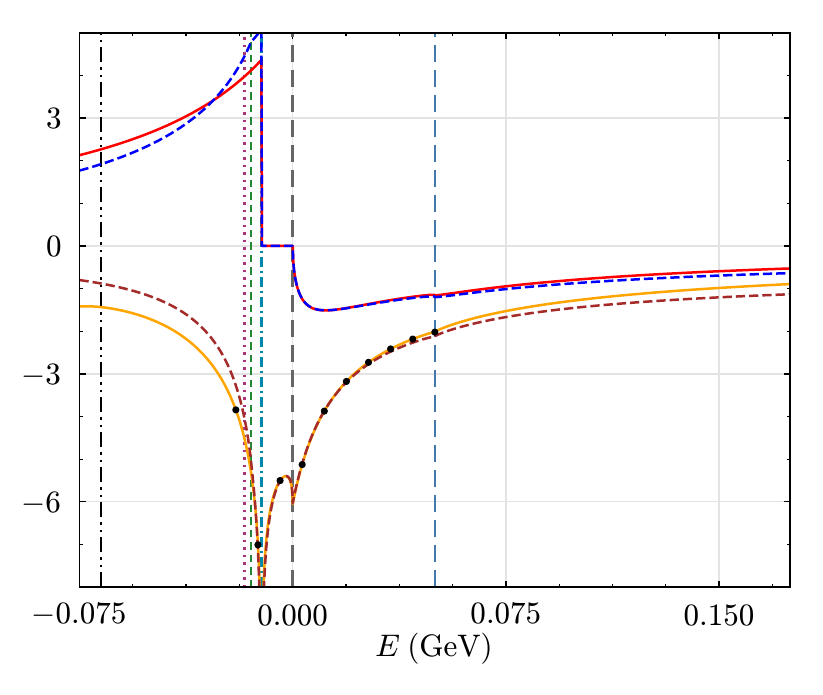}
            \put(70,25){$\mathbb{T}_{22}$}
        \end{overpic}
    \end{minipage}
    \caption{Comparison of the $[0,0]$ coupled-channel parametrization (dashed curves) from Eq.~\eqref{equ:F00} with the LSE solution (solid curves) from Eq.~\eqref{equ:LSE} for each matrix element $\mathbb{T}_{ij}$ of the scattering amplitude. The black dots indicate the fitting points used to determine the six parameters of the $[0,0]$ parametrization from the LSE results. The violet dotted, green short-dashed, and teal dash-dotted vertical lines mark the branch points of the lhc associated with $L_{t,11}$, $L_{t,12}$, and $L_{t,22}$ in Eq.~\eqref{equ:Lintwochannel}, respectively. The black dash-dot-dotted vertical line marks the once-iterated diagonal OPE branch point at $k_1^2=-\mex_{t,11}^2$. The grey and blue long-dashed vertical lines denote the thresholds of the first and second channels.}
    \label{fig:T[00]}
\end{figure}

\subsection{Coupled-channel parameterization vs. LSE}

\begin{figure}[tb]
    \centering
    \begin{minipage}{0.499\linewidth}
        \centering
        \begin{overpic}[width=\linewidth]{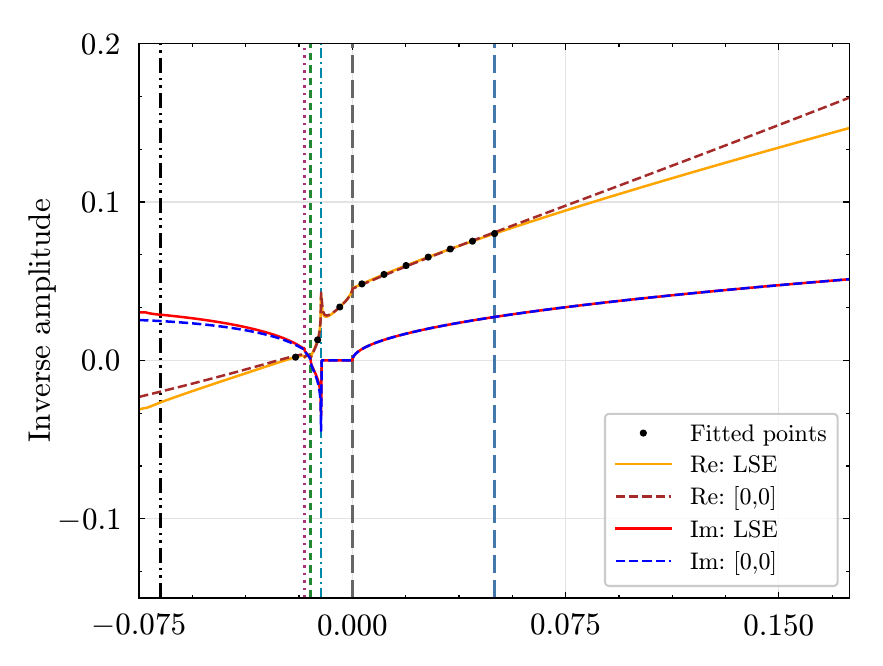}
            \put(25,60){$\mathbb{T}^{-1}_{11}$}  
        \end{overpic}
    \end{minipage}
    \hfill
    \begin{minipage}{0.479\linewidth}
        \centering
        \begin{overpic}[width=\linewidth]{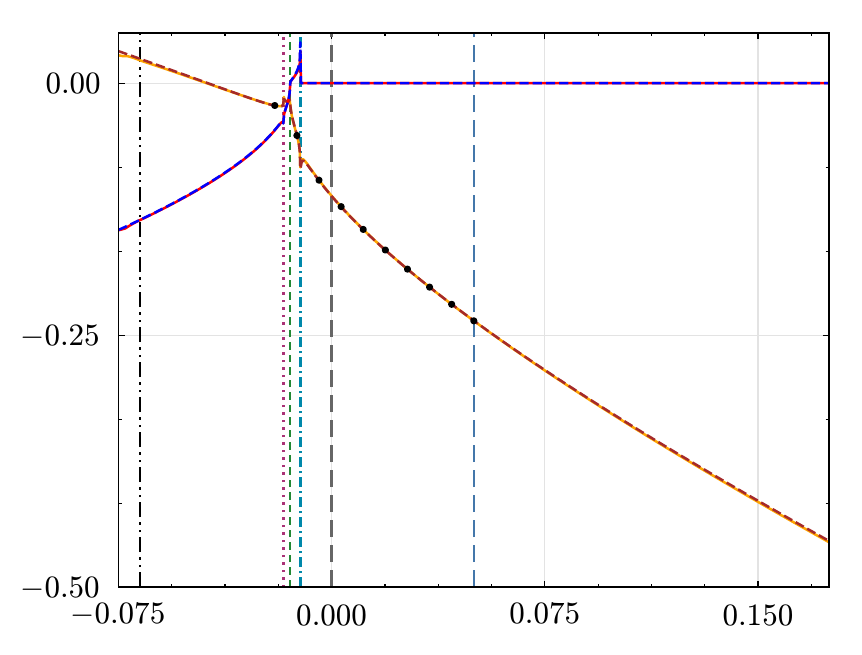}
            \put(22,20){$\mathbb{T}^{-1}_{12}$}
        \end{overpic}
    \end{minipage}
    \\[3mm]
    \begin{minipage}{0.512\linewidth}
        \centering
        \begin{overpic}[width=\linewidth]{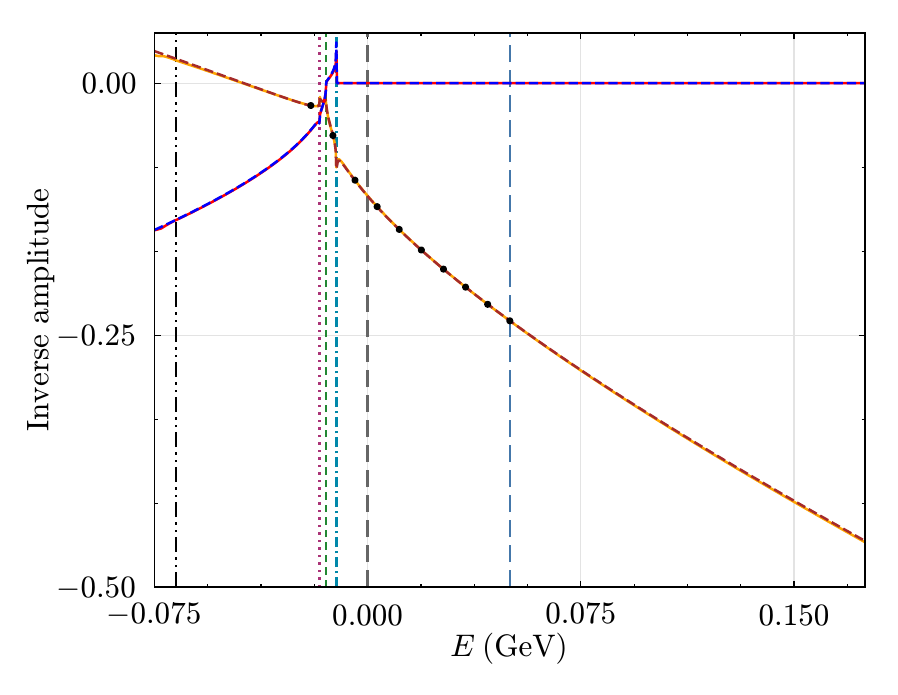}
            \put(25,25){$\mathbb{T}^{-1}_{21}$}
        \end{overpic}
    \end{minipage}
    \hfill
    \begin{minipage}{0.469\linewidth}
        \centering
        \begin{overpic}[width=\linewidth]{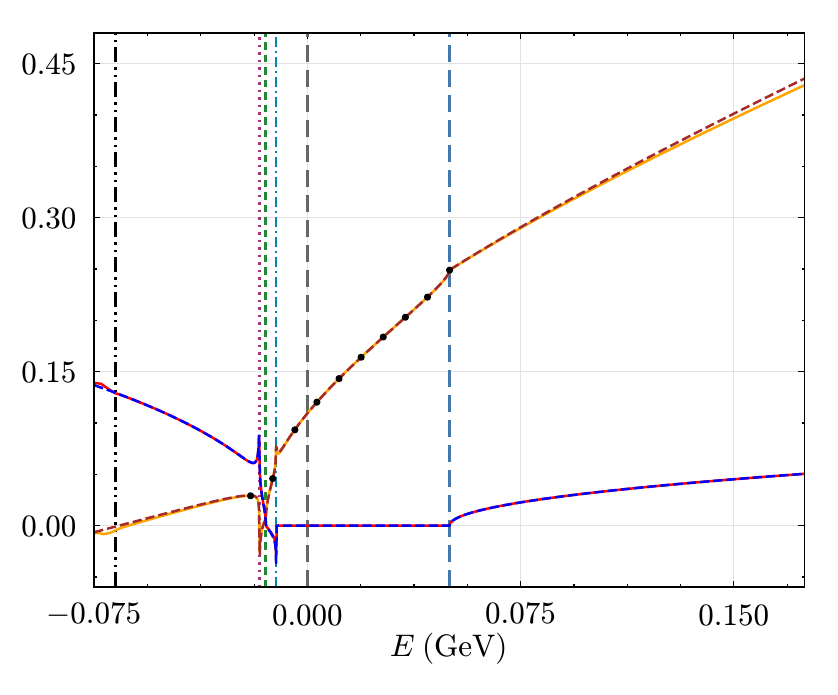}
            \put(20,65){$\mathbb{T}^{-1}_{22}$}
        \end{overpic}
    \end{minipage}
    \caption{Comparison of the inverse amplitude obtained from the $[0,0]$ coupled-channel parametrization (dashed curves) in Eq.~\eqref{equ:F00} with the LSE solution (solid curves) in Eq.~\eqref{equ:LSE}. Each panel displays a different matrix element $\mathbb{T}^{-1}_{ij}$. The notation follows that of Fig.~\ref{fig:T[00]}.}
    \label{fig:inv_T[00]}
\end{figure}

Guided by the constraints in Eqs.~(\ref{constr_three_t}, \ref{equ:commom_for_two}), we adopt a set of particle masses typical of the strong interaction regime, specifically $m_{11}=2$ GeV, $m_{12}=1$ GeV, $m_{21}=1.8$ GeV, $m_{22}=1.25$ GeV, $\mex_{t,11}=0.3$ GeV, $\mex_{t,12}=0.5$ GeV, and $\mex_{t,22}=0.6$ GeV. With these values, Eq.~(\ref{equ:definationofMij}) yield the following branch points: $-M_{12,+}^2=-0.020\ \text{GeV}^2$, $-M_{12,-}^2=-7.720\ \text{GeV}^2$, $-M_{21,+}^2=-0.095\ \text{GeV}^2$, and $-M_{21,-}^2=-8.617\ \text{GeV}^2$. These endpoints play a crucial role in constructing the two-channel parametrization in Eq.~(\ref{equ:Ffortwochannel}). The OPE strength parameters in Eq.~(\ref{equ:VinLSE}) are set to $c_{11}=1.0$, $c_{12}=c_{21}=1.5$, and $c_{22}=0.5$. With these inputs, we solve the LSE~(\ref{equ:LSE}) to obtain the two-channel scattering amplitude matrix.

To reproduce the LSE amplitude, we determine the parameters in Eq.~\eqref{equ:Ffortwochannel} by fitting exclusively to the real parts of the amplitude matrix elements generated from the LSE~(\ref{equ:LSE}).\footnote{Below threshold, the scattering phase shift and inelasticity become complex due to the multivalued nature of functions such as square roots and logarithms, leading to abrupt changes. Directly fitting the amplitude (or its inverse) avoids these complications and validates the parameterization scheme in Eq.~\eqref{equ:Ffortwochannel}.} The fit is performed over the energy range $E \in [-0.02, 0.05]~\text{GeV}$ at ten uniformly distributed points, each assigned equal uncertainty; these points are indicated by black dots in Fig.~\ref{fig:T[00]}. Using only six parameters, namely $c_{11}$, $c_{21}$, $c_{22}$, $\tilde{\mathbb{N}}_{0,11}$, $\tilde{\mathbb{N}}_{0,12}$, and $\tilde{\mathbb{N}}_{0,22}$, the $[0,0]$ approximant in Eq.~\eqref{equ:F00} provides a good description of the LSE amplitude for all matrix elements. In particular, it accurately reproduces the branch points of both the lhc and rhc, as illustrated in Fig.~\ref{fig:T[00]} as they have been built in explicitly. 
Moreover, as shown in Table~\ref{table:Fitresult}, the extracted OPE strength parameters, namely $c_{11}=1.06$, $c_{21}=1.56$, and $c_{22}=0.58$, are in good agreement with the LSE input values. This demonstrates a key advantage of the present method: it enables extraction of coupling strengths, which has been illustrated in a realistic case of nucleon-nucleon scattering in Ref.~\cite{Liu:2026xrk}.
In the plots of Fig.~\ref{fig:T[00]}, we also label the branch point at $k_i^2 = -\mex_{t,11}^2$ due to the once-iteration in Fig.~\ref{fig:once-iterated}. It is clear from the LSE curves, which have all iterations included, that the effects from the once-iteration lhc is almost invisible for the chosen couplings.
Technical subtlety of the LSE with the correct analytic behavior is discussed in Appendix~\ref{app:logintegral}.

\renewcommand{\arraystretch}{1.1}
\begin{table}[tb]
\caption{Extracted parameters from fits to the LSE $\operatorname{Re}[\mathbb{T}]$ or $\operatorname{Re}[\mathbb{T}^{-1}]$, using the $[m,n]$ parameterization of the two-channel $S$-wave scattering amplitude in Eq.~\eqref{equ:Ffortwochannel}. The fit results for rows 1--4 are shown in Figs.~\ref{fig:T[00]},~\ref{fig:inv_T[00]},~\ref{fig:T[01]},~\ref{fig:inv_T[01]}, respectively. The last row lists the  input values for the OPE strength parameters in the LSE.}
\centering
\footnotesize
\begin{tabular}{l|ccc}
\toprule[2pt]
\makebox[10em][l]{Fit scheme} & \makebox[4em][c]{$c_{11}$ } & \makebox[4em][c]{$c_{21}$ } & \makebox[4em][c]{$c_{22}$ } \\
\midrule[0.5pt]
$[0,0]$ for Re$[\mathbb{T}]$ & 1.06  & 1.56 & 0.58 \\
$[0,0]$ for Re$[\mathbb{T}^{-1}]$ & 1.04  & 1.55 & 0.58\\
$[0,1]$ for Re$[\mathbb{T}]$ & 0.99 & 1.49 & 0.50\\
$[0,1]$ for Re$[\mathbb{T}^{-1}]$ &1.00  & 1.50 & 0.50 \\
Input values in LSE & 1.00 &1.50 &0.50  \\
\bottomrule[2pt]
\end{tabular}
\label{table:Fitresult}
\end{table}

\begin{figure}[tb]
    \centering
    \begin{minipage}{0.499\linewidth}
        \centering
        \begin{overpic}[width=\linewidth]{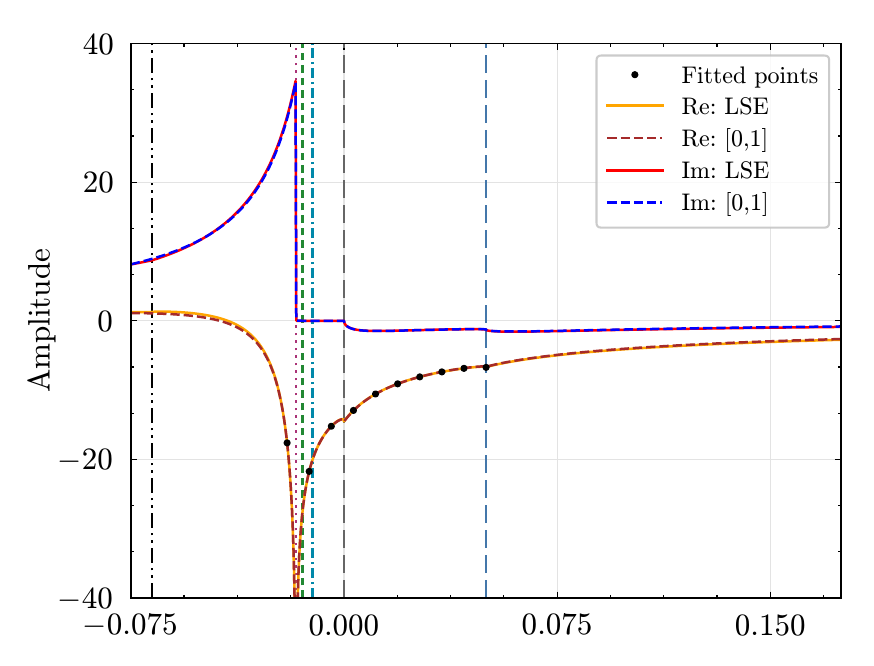}
            \put(70,20){$\mathbb{T}_{11}$}  
        \end{overpic}
    \end{minipage}
    \hfill
    \begin{minipage}{0.479\linewidth}
        \centering
        \begin{overpic}[width=\linewidth]{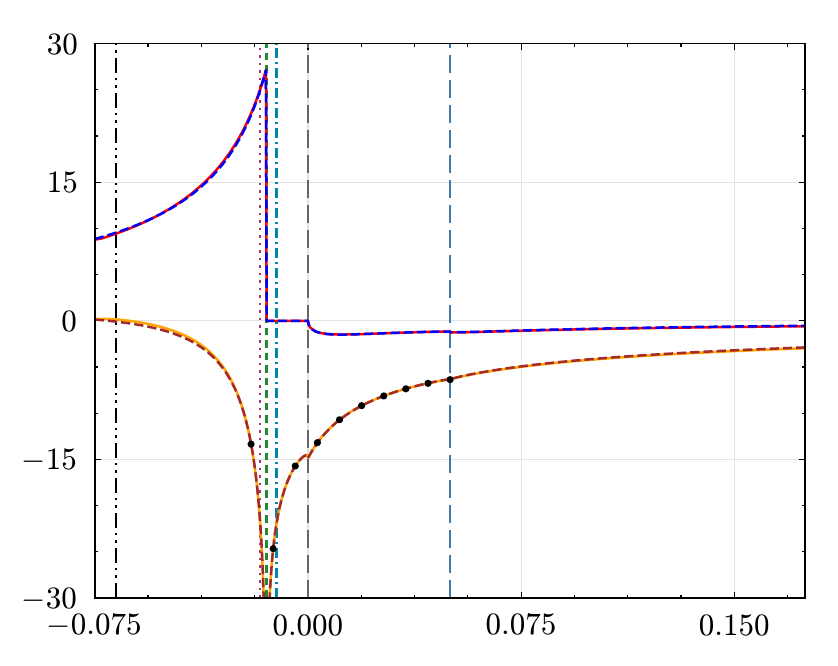}
            \put(70,20){$\mathbb{T}_{12}$}
        \end{overpic}
    \end{minipage}
    \\[3mm]
    \begin{minipage}{0.512\linewidth}
        \centering
        \begin{overpic}[width=\linewidth]{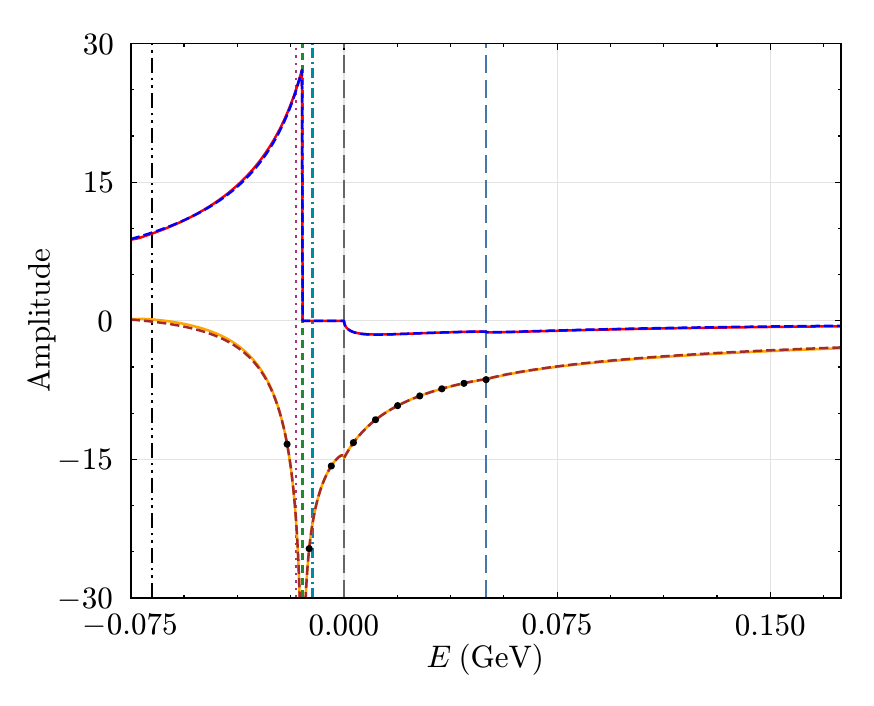}
            \put(70,25){$\mathbb{T}_{21}$}
        \end{overpic}
    \end{minipage}
    \hfill
    \begin{minipage}{0.469\linewidth}
        \centering
        \begin{overpic}[width=\linewidth]{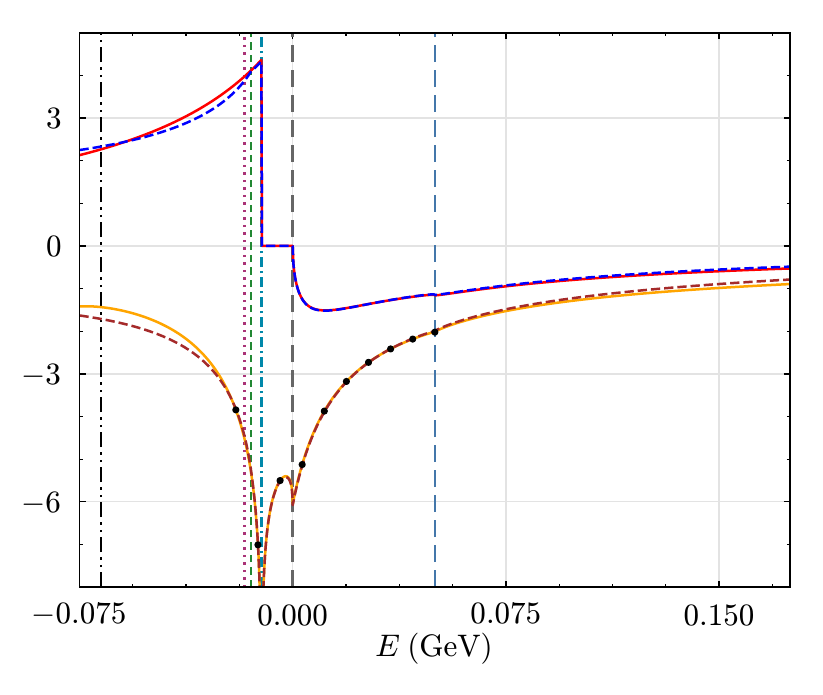}
            \put(70,25){$\mathbb{T}_{22}$}
        \end{overpic}
    \end{minipage}
    \caption{Comparison of the $[0,1]$ coupled-channel parametrization (dashed curves) with the LSE solution (solid curves) for $\mathbb{T}_{ij}$. Notation follows Fig.~\ref{fig:T[00]}.}
    \label{fig:T[01]}
\end{figure}

Alternatively, we also determine the parameters of the $[0,0]$ approximant in Eq.~\eqref{equ:F00} by fitting to the real parts of the inverse amplitude $\mathbb{T}^{-1}$, directly related to the phase shifts and inelasticities, generated from the LSE~\eqref{equ:F00}. The corresponding fit result is shown in Fig.~\ref{fig:inv_T[00]}. The $[0,0]$ approximant again provides a good description of all matrix elements. Moreover, as shown in Table~\ref{table:Fitresult}, the extracted OPE coupling strength parameters, namely $c_{11}=1.04$, $c_{21}=1.55$, $c_{22}=0.58$, remain in good agreement with the LSE input values.

\begin{figure}[tb]
    \centering
    \begin{minipage}{0.499\linewidth}
        \centering
        \begin{overpic}[width=\linewidth]{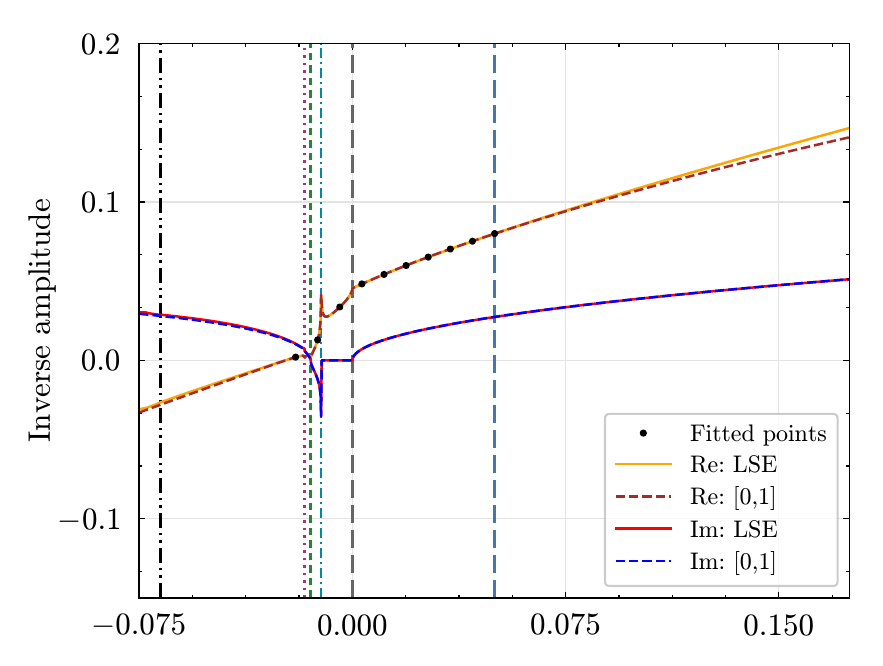}
            \put(25,60){$\mathbb{T}^{-1}_{11}$}  
        \end{overpic}
    \end{minipage}
    \hfill
    \begin{minipage}{0.479\linewidth}
        \centering
        \begin{overpic}[width=\linewidth]{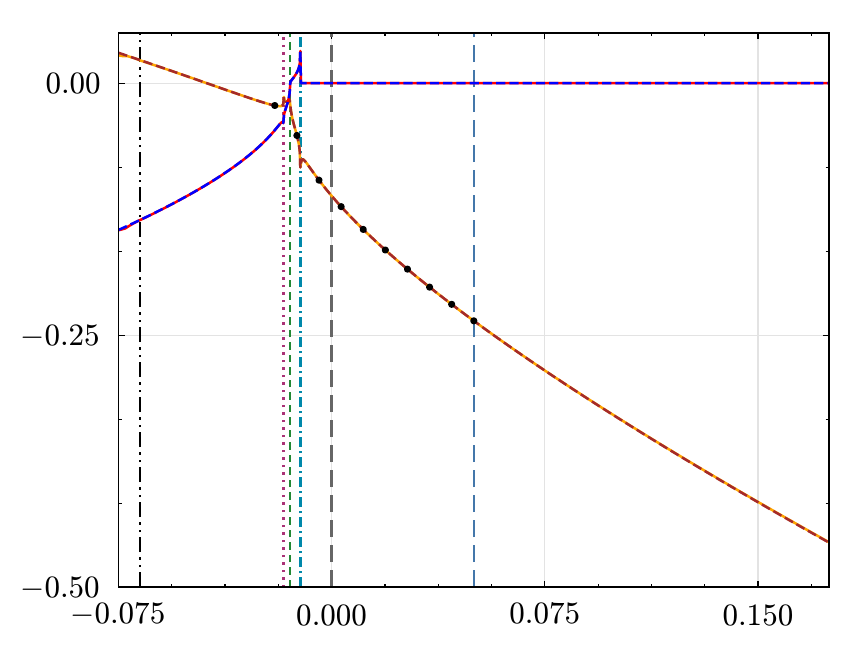}
            \put(22,20){$\mathbb{T}^{-1}_{12}$}
        \end{overpic}
    \end{minipage}
    \\[3mm]
    \begin{minipage}{0.512\linewidth}
        \centering
        \begin{overpic}[width=\linewidth]{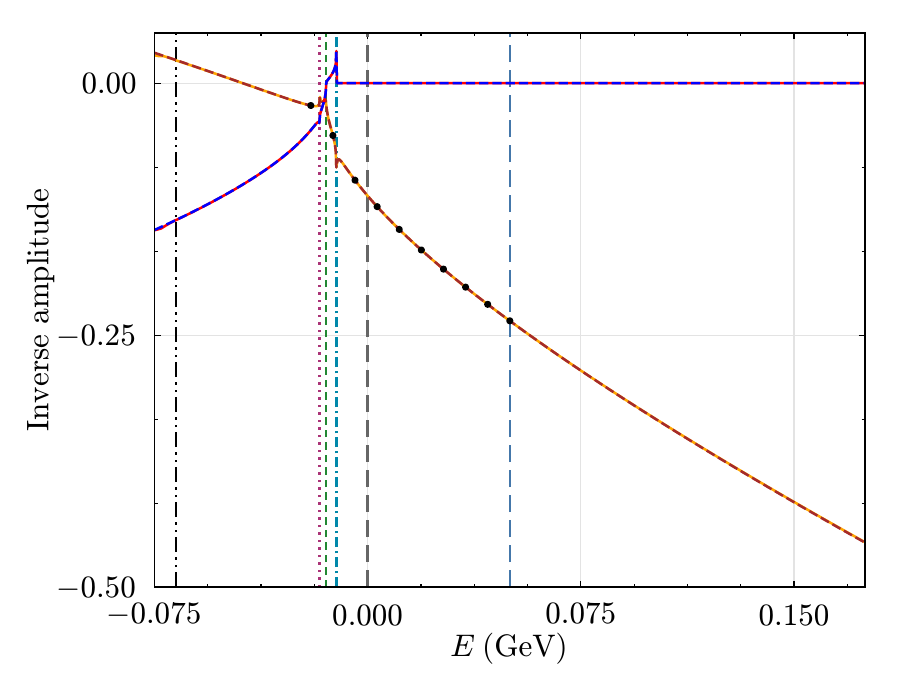}
            \put(25,25){$\mathbb{T}^{-1}_{21}$}
        \end{overpic}
    \end{minipage}
    \hfill
    \begin{minipage}{0.469\linewidth}
        \centering
        \begin{overpic}[width=\linewidth]{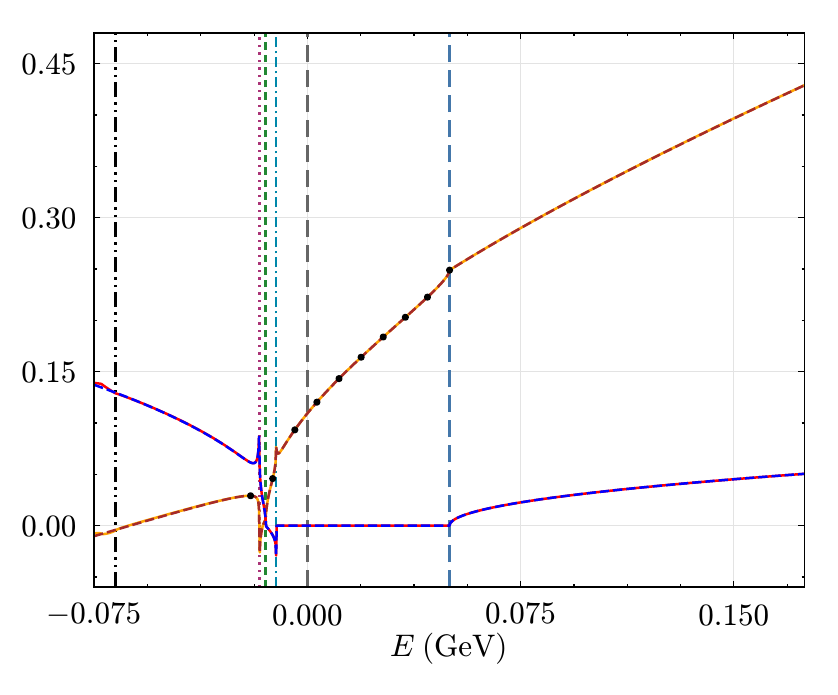}
            \put(20,65){$\mathbb{T}^{-1}_{22}$}
        \end{overpic}
    \end{minipage}
    \caption{Comparison of the $[0,1]$ coupled-channel parametrization (dashed curves) with the LSE solution (solid curves) for $\mathbb{T}^{-1}_{ij}$. Notation follows Fig.~\ref{fig:T[00]}.}
    \label{fig:inv_T[01]}
\end{figure}

The approximation of the parametrization to the LSE solution can be improved by considering higher order approximants.
At the $[0,1]$ order, which introduces three more parameters $\tilde{\mathbb{D}}_{1,11}$, $\tilde{\mathbb{D}}_{1,21}$, and $\tilde{\mathbb{D}}_{1,22}$, we compare in Figs.~\ref{fig:T[01]} and~\ref{fig:inv_T[01]} the resulting amplitude and its inverse with the LSE solution. Relative to the $[0,0]$ results, the $[0,1]$ approximant reproduces the LSE amplitude more accurately for all matrix elements, particularly in energy regions slightly above the threshold. Furthermore, as shown in Table~\ref{table:Fitresult}, the extracted OPE strength parameters are in perfect agreement with the LSE input values. These results confirm the robustness and effectiveness of the proposed parametrization.

\section{summary}\label{sec:summary}

We have systematically extended the generalized ERE, incorporating the lhc generated by OPE, to a coupled-channel framework. We have performed a comprehensive analysis of the singularity structure of the $S$-wave coupled-channel OPE potential under various mass configurations. Based on this analysis and employing the $N/D$ method, we have derived an analytical parametrization of the coupled-channel scattering amplitude that explicitly includes the lhc from the tree-level potential for multiple channels.
Our new parametrization adds only coupling constants associated with the exchanged particle and the scattering particles compared to the conventional coupled-channel ERE.

Focusing on the two-channel case with the typical mass configuration, we benchmark this coupled-channel parameterization against the numerical solution of the LSE that incorporates the OPE lhcs at different expansion orders. 
Our results demonstrate that the coupled-channel parameterization successfully reproduces the full LSE amplitudes in the near-threshold region beyond the OPE lhc, even though we fit only the real parts of these quantities. 
Even at the lowest $[0,0]$ order with only six parameters, the parameterization can successfully reproduce the LSE amplitudes and allows for an extraction of the OPE coupling strengths. As the expansion order increases, the agreement improves. 
These findings establish the robustness and effectiveness of the proposed framework.

\begin{acknowledgements}
We are grateful to Xiong-Hui Cao, Yong-Hui Lin, Mao-Jun Yan, and Zhi Yang for useful discussions. This work is supported in part by the National Natural Science Foundation of China (NSFC) under Grants No.~12675095, 12547111, 12125507, 12361141819, 12447101, and 12405100; by the National Key R\&D Program of China under Grant No.~2023YFA1606703; and by the Chinese Academy of Sciences under Grant No.~YSBR-101.
\end{acknowledgements}

\begin{appendix}
    
\section{Classification of the lhc in $L_{t,ij}(k_j^2)$}\label{sec:appendixofsingularity}
In this appendix, we present a detailed analysis of the singularities of $L_{t,ij}(k_j^2)$ ($i>j$), as given by Eq.~(\ref{equLt_nonreal}), on the real axis of the complex $k_j^2$-plane. Singularities off the real axis are discussed separately in Appendix~\ref{sec:appendixofcircularcut}.

\begin{figure}[tb]
\centering
\begin{tikzpicture}[scale=0.7]
\begin{scope}
\draw[-latex,thick] (-4,0) -- (0,0) node [below=6pt, right,black] {0}-- (1.1,0) node[right] {${\rm Re}[k_j^2]$};
\draw[-latex,thick] (0,-2) -- (0,2.1) node[above] {${\rm Im}[k_j^2]$};
\draw[blue,very thick] (-3,0) node[below] {$-M_{ji,-}^2$} -- (-1,0) node[below] {$-M_{ji,+}^2$};
\fill[fill=blue] (-1,0) circle[radius=2pt];
\fill[fill=blue] (-3,0) circle[radius=2pt];
\node[blue] at (-2.0,0.75) {$-\frac{\pi}{4k_ik_j}$};
\node[red] at (-1.5,3.5) {Case 1: $\mex_{t,ij}^2>A_{ij}$};
\end{scope}

\begin{scope}[xshift=9cm]
\draw[-latex,thick] (-4,0) -- (0,0) node [below=6pt, right,blue] {0}-- (1.1,0) node[right] {${\rm Re}[k_j^2]$};
\draw[-latex,thick] (0,-2) -- (0,2.1) node[above] {${\rm Im}[k_j^2]$};
\draw[blue,very thick] (-3,0) node[below] {$-M_{ji,-}^2$} -- (0,0);
\fill[fill=blue] (0,0) circle[radius=2pt];
\fill[fill=blue] (-3,0) circle[radius=2pt];
\node[blue] at (-1.5,0.75) {$-\frac{\pi}{4k_ik_j}$};
\node[red] at (-1.5,3.5) {Case 2: $\mex_{t,ij}^2=A_{ij}$};
\end{scope}

\begin{scope}[xshift=-1.0cm,yshift=-8cm]
\draw[-latex,thick] (-4,0) -- (0,0) node [below=6pt, right,purple] {0}-- (1.1,0) node[right] {${\rm Re}[k_j^2]$};
\draw[-latex,thick] (0,-2) -- (0,2.1) node[above] {${\rm Im}[k_j^2]$};
\draw[blue,very thick] (-3.5,0) node[below] {$-M_{ji,-2}^2$} -- (-1.5,0) node[below] {$-M_{ji,-1}^2$};
\draw[purple,very thick] (-1.5,0) -- (0,0);
\fill[fill=purple] (0,0) circle[radius=2pt];
\fill[fill=blue] (-1.5,0) circle[radius=2pt];
\fill[fill=blue] (-3.5,0) circle[radius=2pt];
\node[blue] at (-2.75,0.75) {$-\frac{\pi}{4k_ik_j}$};
\node[purple] at (-0.75,0.75) {$-\frac{\pi}{2k_ik_j}$};
\node[red] at (-2.0,4.5) {Case 3: $\mex_{t,ij}^2\in (|m_{i1}-m_{j1}|^2,A_{ij}),\ \frac{m_{i1}}{m_{j1}}\geq\sqrt{\frac{\mu_i}{\mu_j}}$};
\node[red] at (-2.4,3.5) {or $\mex_{t,ij}^2\in(B_{ij},A_{ij}),\ \frac{m_{i1}}{m_{j1}}<\sqrt{\frac{\mu_i}{\mu_j}}$};
\end{scope}

\begin{scope}[xshift=10.0cm,yshift=-8cm]
\draw[-latex,thick] (-4,0) -- (0,0) node [below=6pt, right,blue] {0}-- (1.1,0) node[right] {${\rm Re}[k_j^2]$};
\draw[-latex,thick] (0,-2) -- (0,2.1) node[above] {${\rm Im}[k_j^2]$};
\draw[blue,very thick] (-3.5,0) node[below] {$\frac{2m_{i1}^2\mu_j^2(m_{j\rm th}-m_{i\rm th})}{m_{j1}^2\mu_i-m_{i1}^2\mu_j}$} -- (0,0);
\fill[fill=blue] (0,0) circle[radius=2pt];
\fill[fill=blue] (-3.5,0) circle[radius=2pt];
\node[blue] at (-2.0,0.75) {$-\frac{\pi}{2k_ik_j}$};
\node[red] at (-1.5,4.0) {Case 4: $\mex_{t,ij}^2\in(|m_{i1}-m_{j1}|^2,B_{ij}],\frac{m_{i1}}{m_{j1}}<\sqrt{\frac{\mu_i}{\mu_j}}$};
\node[red] at (-2.5,1.75) {Circular cut appears};
\end{scope}
\end{tikzpicture}
\caption{Singularity structure of $L_{t,ij}(k_j^2)$ ($i > j$) defined in Eq.~\eqref{equLt_nonreal} on the complex $k_j^2$-plane for the scenarios listed in Eq.~\eqref{8Cases}. The values $-\pi/(4k_i k_j)$ and $-\pi/(2k_i k_j)$ represent the imaginary part of $L_{t,ij}(k_j^2)$ along the corresponding cut, i.e., $\operatorname{Im} L_{t,ij}(k_j^2)$.}
    \label{fig:Singularityfortwochannel_all}
\end{figure}
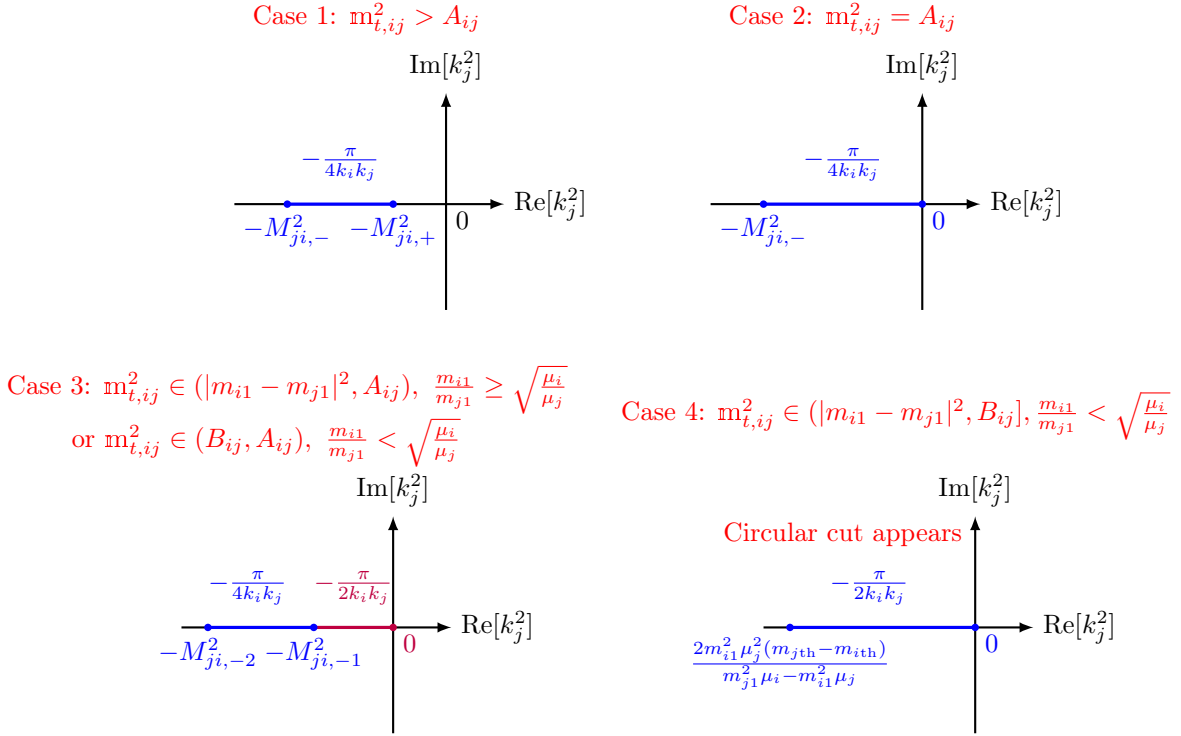

We first note that $L_{t,ij}(k_j^2)$ is real for $k_j^2\geq0$.\footnote{For $k_j^2\in\left(0,2\mu_j(m_{1\rm th}-m_{j\rm th})\right)$, the combination $\left[\log f_{-,ij}(k_j^2)-\log f_{+,ij}(k_j^2)\right]$ is purely imaginary, and the prefactor $1/(4k_i k_j)$ is also purely imaginary in this region. Their product in Eq.~(\ref{equLt_nonreal}) therefore yields a real $L_{t,ij}(k_j^2)$.} Hence,
\begin{align}
    {\rm Im}\,L_{t,ij}(k_j^2)=0\ ,\qquad\text{ for }k_j^2\geq0\ .
\end{align}
Consequently, it suffices to examine the negative real axis $k_j^2<0$, where the lhc arises when the arguments of the logarithms become negative, i.e., $f_{\pm,ij}(k_j^2)<0$. To proceed, we note the following properties of $f_{\pm,ij}(k_j^2)$:
\begin{align}
    f_{\pm,ij}(k_j^2=0)&=\mex_{t,ij}^2-A_{ij}\ ,\\
    \frac{{\rm d}f_{\pm,ij}(k_j^2)}{{\rm d}k_j^2}&=\frac{m_{i1}}{m_{j1}}+\frac{m_{j1}}{m_{i1}}\frac{\mu_i}{\mu_j}\pm \left( \frac{k_i}{k_j}+\frac{k_j}{k_i}\frac{\mu_i}{\mu_j} \right) \ ,\\
    f_{+,ij}(k_j^2=-\infty)&=-\infty\ ,\\
    f_{-,ij}(k_j^2=-\infty)&= \begin{cases}
   -\infty\ ,& \text{  for }\frac{m_{i1}}{m_{j1}}\neq\sqrt{\frac{\mu_i}{\mu_j}}\ , \\
   \mex_{t,ij}^2-(m_{i1}-m_{j1})^2\ ,& \text{  for }\frac{m_{i1}}{m_{j1}}=\sqrt{\frac{\mu_i}{\mu_j}}\ ,
   \end{cases}\\
   {\rm Im}\,L_{t,ij}(k_j^2)&=\frac{{\rm Im}\left[\log f_{-,ij}(k_j^2+i\epsilon)\right]-{\rm Im}\left[\log f_{+,ij}(k_j^2+i\epsilon)\right]}{4k_ik_j}\ ,\\
   {\rm Im}\left[ \log f_{\pm,ij}(k_j^2+i\epsilon)\right]&=
   \begin{cases}
   0\ ,& \text{  for }f_{\pm,ij}(k_j^2)>0\ , \\
   \pi\ ,& \text{  for }f_{\pm,ij}(k_j^2)<0\ ,\frac{{\rm d}f_{\pm,ij}(k_j^2)}{{\rm d}k_j^2}>0\ ,\\
   -\pi\ ,& \text{  for }f_{\pm,ij}(k_j^2)<0\ ,\frac{{\rm d}f_{\pm,ij}(k_j^2)}{{\rm d}k_j^2}<0\ ,
   \end{cases}
\end{align}
where $\epsilon\to 0^+$. Analyzing these conditions leads to four distinct scenarios as shown in Fig.~\ref{fig:Singularityfortwochannel_all}, classified as follows:
\begin{equation}\label{8Cases}
\begin{aligned}
   \text{Case 1: }& \mex_{t,ij}^2>A_{ij},\quad{\rm Im}\ L_{t,ij}(k_j^2)=-\frac{\pi}{4k_ik_j}\ \text{  for }k_j^2\in[-M_{ji,-}^2,-M_{ji,+}^2]\ ;\\
   \text{Case 2: }& \mex_{t,ij}^2=A_{ij},\quad{\rm Im}\ L_{t,ij}(k_j^2)=-\frac{\pi}{4k_ik_j}\  \text{  for }k_j^2\in[-M_{ji,-}^2,0]\ ;\\
   \text{Case 3: }& \mex_{t,ij}^2\in (|m_{i1}-m_{j1}|^2,A_{ij}),\ \frac{m_{i1}}{m_{j1}}\geq\sqrt{\frac{\mu_i}{\mu_j}}\quad\text{or}\quad\mex_{t,ij}^2\in(B_{ij},A_{ij}),\ \frac{m_{i1}}{m_{j1}}<\sqrt{\frac{\mu_i}{\mu_j}},\\
   &{\rm Im}\ L_{t,ij}(k_j^2)=
   \begin{cases}
   -\frac{\pi}{4k_ik_j}\ ,& \text{  for }k_j^2\in[-M_{ji,-2}^2,-M_{ji,-1}^2]\ , \\
   -\frac{\pi}{2k_ik_j}\ ,& \text{  for }k_j^2\in[-M_{ji,-1}^2,0]\ ;
   \end{cases}\\
   \text{Case 4: }& \mex_{t,ij}^2\in(|m_{i1}-m_{j1}|^2,B_{ij}],\ \frac{m_{i1}}{m_{j1}}<\sqrt{\frac{\mu_i}{\mu_j}},\\
   &{\rm Im}\ L_{t,ij}(k_j^2)=-\frac{\pi}{2k_ik_j}\  \text{ for }k_j^2\in\left[\frac{2m_{i1}^2\mu_j^2(m_{j\rm th}-m_{i\rm th})}{m_{j1}^2\mu_i-m_{i1}^2\mu_j},0\right], \text{ circular cut appears.}
\end{aligned}
\end{equation}
The auxiliary quantities $A_{ij}$ and $B_{ij}$ appearing above are defined as
\begin{align}
    A_{ij}&=(m_{i1}-m_{j1})^2+2(m_{i\rm th}-m_{j\rm th})\frac{m_{j1}}{m_{i1}}\mu_i\ ,\\
    B_{ij}&=(m_{i1}-m_{j1})^2+2(m_{i\rm th}-m_{j\rm th})\left(\frac{m_{j1}}{m_{i1}}\mu_i-\frac{m_{i1}}{m_{j1}}\mu_j \right) ,
\end{align}
and the various $-M_{ji,\pm}^2$ denote the roots of $f_{\pm,ij}(k_j^2)=0$ under the corresponding mass configurations, i.e.,
\begin{equation}
\begin{aligned}\label{equ:Mij}
    &f_{\pm,ij}(k_j^2=-M_{ji,\pm}^2)=0\ ,& &\text{for Cases 1 and 2\ ,}\\
    &f_{-,ij}(k_j^2=-M_{ji,-1}^2)=f_{-,ij}(k_j^2=-M_{ji,-2}^2)=0\ , & &\text{for Case 3}\ ,
\end{aligned}
\end{equation}
where $-M_{ji,-1}^2 > -M_{ji,-2}^2$, and the subscripts $-1$ and $-2$ label the two roots of $f_{-,ij}(k_j^2)$. The analytic expression for $-M_{ji,\pm}^2$ is rather involved and is not presented here. In practice, once the particle masses are specified, these values can be obtained straightforwardly by numerically solving Eq.~\eqref{equ:Mij}.

In light of the constraints in Eqs.~(\ref{constr_three_t}, \ref{relationkikj}), Case 1 represents the most typical scenario, corresponding to the generic conditions $\mex_{t,ij}^2 > A_{ij}$. Notably, apart from Case 4, which involves a circular cut off the real axis, all other cases exhibit branch cuts located exclusively on the negative real axis of the complex $k_j^2$-plane, i.e., the conventional lhc. A detailed discussion of the circular cut is provided in Appendix~\ref{sec:appendixofcircularcut}. Moreover, for Cases 1--3, when $m_{i1}/m_{j1} = \sqrt{\mu_i/\mu_j}$, the lhc extends to negative infinity; the single $t$-channel case in Eq.~\eqref{equ:ImLtii} corresponds precisely to this situation.

A key observation is that in Cases 2--4 in Fig.~\ref{fig:Singularityfortwochannel_all}, where the exchanged particle mass is so small that it nearly reaches the point where three-body kinematics would emerge at the two-body threshold, the lhc branch point approaches the lowest threshold value. This behavior resembles that of a massless-particle propagator (e.g., a photon), 
indicating that under specific mass configurations, a massive system can develop an lhc structure analogous to the massless-particle-exchange situation.

An alternative and more systematic way for performing the above analysis can be done using the discontinuity calculus method developed in Ref.~\cite{Jing:2025qmi}.

\section{Circular cut in $L_{t,ij}(k_j^2)$}\label{sec:appendixofcircularcut}

\begin{figure}[tb]
    \centering
    \begin{tikzpicture}[scale=0.7]
    
    \draw[-latex,thick] (-4,0) -- (0,0) node [below=6pt, left,black] {0}-- (4.1,0) node[right] {${\rm Re}[k_j^2]$};
    \draw[-latex,thick] (0,-4) -- (0,4.1) node[above] {${\rm Im}[k_j^2]$};
    \draw[blue,dashed, thick] (-0.5, -2) arc[start angle=-110, end angle=110, x radius=3, y radius=2.08];
    \draw[blue, very thick] (-0.5, -2) arc[start angle=-110, end angle=-250, x radius=3, y radius=2.08];
    \draw[red,thick,dotted] (-0.5,-3.5) -- (-0.5,3.5);
    \fill[fill=blue] (0.5,0) circle[radius=2pt];
    \node[blue] at (0.75,-0.5) {$-\frac{b_{ij}}{2}$};
    \node[cyan] at (-4.5,-0.75) {$\frac{b_{ij}}{2}\left( \sqrt{\frac{a_{ij}^2}{a_{ij}^2-4}}-1 \right)$};
    \node[red] at (-1.75,-3) {$\frac{2b_{ij} - a_{ij}c_{ij}}{a_{ij}^2-4}$};
    \fill[fill=cyan] (-2.48,0) circle[radius=2pt];

    \end{tikzpicture}
    \caption{Circular cut in the complex $k_j^2$-plane as given by Eq.~(\ref{Circularcut}). The left segment of the ellipse (solid blue line) represents the actual cut, while the dashed part is excluded by the inequality constraint.}
    \label{fig:circularcut}
\end{figure}
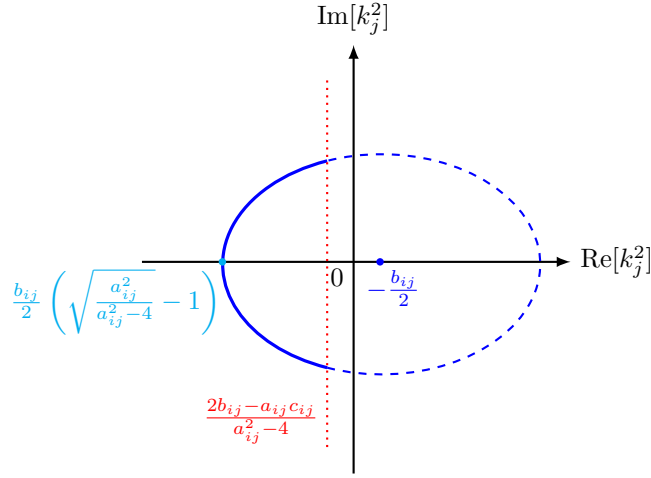

The singularities of $L_{t,ij}(k_j^2)$ arise from the branch cut of the logarithmic function in Eq.~(\ref{equLt_nonreal}) along the negative real axis. Specifically, they occur in regions of the complex $k_j^2$-plane where the arguments $f_{\pm,ij}(k_j^2)$ satisfy
\begin{align}
    {\rm Im}\,f_{\pm,ij}(k_j^2)=0,\quad {\rm Re}\,f_{\pm,ij}(k_j^2)<0\ .
\end{align}
Solving these conditions reveals that, in addition to the lhc on the negative real axis (${\rm Im}\,k_j^2=0$ and ${\rm Re}\,k_j^2<0$, discussed in Appendix~\ref{sec:appendixofsingularity}), there exists a circular cut in the complex $k_j^2$-plane characterized by
\begin{align}
 {\rm Im}\,k_j^2\neq0,\qquad
   \begin{cases}
    \frac{\left({\rm Re}\,k_j^2+b_{ij}/2\right)^2}{a_{ij}^2b_{ij}^2/[4(a_{ij}^2-4)]} + \frac{({\rm Im}\,k_j^2)^2}{b_{ij}^2/(a_{ij}^2-4)}=1\ ,\\
    {\rm Re}\,k_j^2 <\frac{2b_{ij} - a_{ij}c_{ij}}{a_{ij}^2-4}\ ,
   \end{cases}\label{Circularcut}
\end{align}
where the parameters are defined as
\begin{align}
    a_{ij}&=\sqrt{\frac{\mu_j}{\mu_i}}\frac{m_{i1}}{m_{j1}}+\sqrt{\frac{\mu_i}{\mu_j}}\frac{m_{j1}}{m_{i1}}\ ,\\
    b_{ij}&=2\mu_j(m_{j\rm th}-m_{i\rm th})\ ,\\
    c_{ij}&=\sqrt{\frac{\mu_j}{\mu_i}}\left[2(m_{j\rm th}-m_{i\rm th})\frac{m_{j1}}{m_{i1}}\mu_i+\mex_{t,ij}^2 - (m_{j1}-m_{i1})^2 \right]\ .
\end{align}
Eq.~(\ref{Circularcut}) describes an ellipse in the complex $k_j^2$-plane, centered at $(-b_{ij}/2, 0)$, with semi-major axis $a_{ij}\sqrt{b_{ij}^2/[4(a_{ij}^2-4)]}$ along the real direction and semi-minor axis $\sqrt{b_{ij}^2/(a_{ij}^2-4)}$ along the imaginary direction. Owing to the inequality constraint in Eq.~(\ref{Circularcut}), however, only the left segment of this ellipse actually contributes to the cut, as depicted in Fig.~\ref{fig:circularcut}.

We do not consider this circular cut in our analysis for the following reasons: its appearance is rare, occurring only under specific mass configurations, and, even when it is present, the cut is incomplete, comprising only the left portion of the ellipse. These features indicate that a proper treatment would require incorporating three-body effects.\footnote{It can be shown that the complete ellipse emerges when $\mex_{t,ij} \le |m_{i1}-m_{j1}|$. Under this condition, Eq.~(\ref{constr_three_t}) implies that three-body kinematics necessarily enters the physical region.} Such an analysis is technically involved and is left to future work.

\section{Useful integrals for constructing $\mathbb{D}^{\rm R}$}\label{sec:calofdR}
The matrix $\mathbb{D}^{\rm R}$ is constructed from integrals that are evaluated using the kinematic relation in Eq.~(\ref{relationkikj}):
\begin{equation}
{k_j^{(\prime)}}^2 = \frac{\mu_j}{\mu_i}{k_i^{(\prime)}}^2 + 2\mu_j(m_{i\rm th}-m_{j\rm th}).
\end{equation}
The basic integrals encountered are
\begin{align}
\int\frac{{\rm d}k'^2}{k'(k'^2-k^2)} &= -\frac{1}{k}\log\left( \frac{ik + i\sqrt{k'^2}}{ik - i\sqrt{k'^2}}\right),\label{int1}\\
\int\frac{{\rm d}k_j'^2}{k_i'(k_j'^2-k_j^2)} &= -\frac{1}{k_i}\log\left[ \frac{ik_i + i\sqrt{k_i'^2}}{ik_i - i\sqrt{k_i'^2}} \right],\label{int2}\\
\int\frac{{\rm d}k_i'^2}{k_j'(k_i'^2-k_i^2)} &= -\frac{1}{k_j}\log\left[ \frac{ik_j + i\sqrt{k_j'^2}}{ik_j - i\sqrt{k_j'^2}} \right].\label{int3}
\end{align}
These integrals provide the building blocks for the explicit expressions of $\mathbb{D}^{\rm R}$ used in the main text.

\section{Continuation of the integral in LSE}
\label{app:logintegral}

In this appendix, we discuss some technical subtleties in solving the coupled-channel LSE. After performing the angular integration of the LSE, one obtains the logarithmic functions from the OPE potential as given in Eq.~\eqref{equ:VinLSE}.
These are part of the integrand of the radial integration in Eq.~\eqref{equ:LSE}. 
For complex momenta, we rewrite Eq.~\eqref{equ:VinLSE} as
\begin{align}
    \mathbb{V}_{ij}(p_i,q_j)&=\frac{c_{ij}}{4p_iq_j}\left[\ell_{ij,-}(p_i,q_j)-\ell_{ij,+}(p_i,q_j)\right],
 \qquad \ell_{ij,\pm}\equiv \Log w_{ij,\pm}+2\pi i \, n_{ij,\pm}, \notag \\
 w_{ij,\pm}(p_i,q_j)&\equiv\mex_{t,ij}^2-(m_{i1}-m_{j1})^2
 +\left(\sqrt{\frac{m_{j1}}{m_{i1}}}p_i\pm\sqrt{\frac{m_{i1}}{m_{j1}}}q_j\right)^2,
 \label{eq:contour_logs}
\end{align}
where we use $\Log$ to denote the principal logarithm, and the integers $n_{ij,\pm}$ are fixed by continuous continuation from real scattering momenta $p_i$, where both logarithmic arguments are positive (at this point $n_{ij,\pm}=0$). When the momenta $p_i$ are analytically continued, the integration contour will be deformed and $n_{ij,\pm}$ increases (decreases) by one whenever $w_{ij,\pm}$ crosses the negative real axis in its complex plane from above to below (from below to above).
It is important to notice that naively taking the principal logarithm of their ratio as written in Eq.~\eqref{equ:VinLSE} in numerical calculations at each energy may lead to an incorrect result that breaks this analytic continuation.

Here we use the following basic example to illustrate the interplay of various singularities and the prescriptions for a correct analytic continuation: 
\begin{align}
    J(p;m)=\int_{C_p}\frac{q\,dq}{p^2-q^2 + i\epsilon}
    \bigl\{\log[w_-(q)] - \log[w_+(q)]\bigr\}, \qquad \text{with } w_\pm \equiv (q\pm p)^2+m^2,
    \label{eq:contour_triangle}
\end{align}
where $C_p$ denotes the contour of the integration. The two logarithms are continued as in Eq.~\eqref{eq:contour_logs} and will be denoted as $\ell_-$ and $\ell_+$, respectively. 
This integral is part of the LSE and also enters into the calculation of a triangle diagram (see also the analysis of triangle singularity using a similar strategy in Ref.~\cite{Bayar:2016ftu}). Here $p$ denotes the on-shell momentum, $q$ denotes the integration momentum, and $m$ is the mass of the $t$-channel exchanged particle.
The poles of the integrand are located at $q=\pm \sqrt{p^2 + i\epsilon}$, and there are four logarithmic branch points located at $q=\pm p\pm im$.

\begin{figure}[tb]
    \centering
    \includegraphics[width=0.9\linewidth]{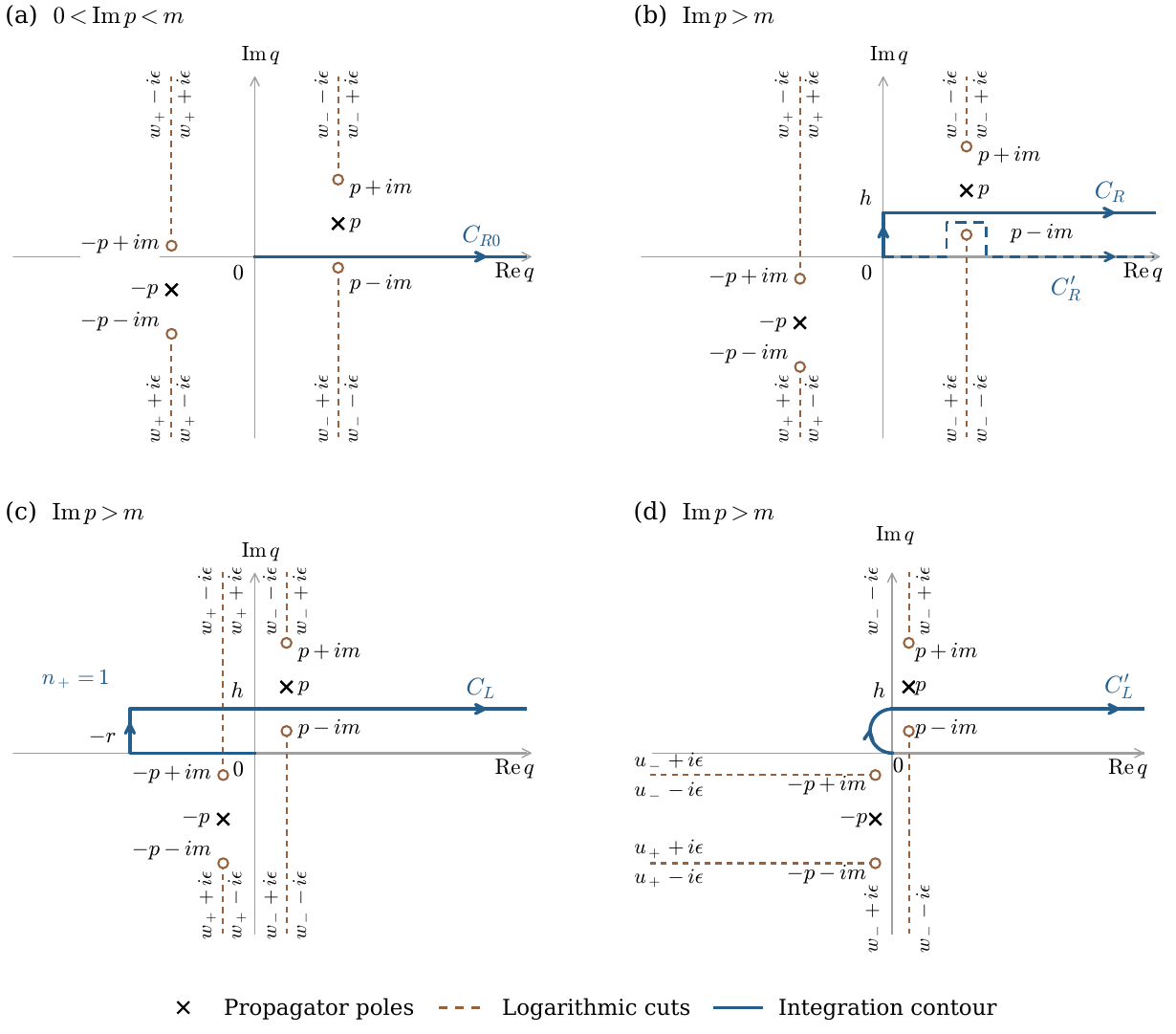}
    \caption{Illustration of the singularities of the integrand in Eq.~\eqref{eq:contour_triangle} and various choices of contour. 
    The crosses and open circles denote the propagator poles and logarithmic branch points, respectively. The brown dashed lines represent the logarithmic cuts, where the two edges of the cuts are labeled by their corresponding values in the $w_\pm$ plane $w_{\pm}\pm i\epsilon$ [and also $u_\pm \pm i\epsilon$ for the horizontal cuts in panel (d)]. 
    (a) for $0<\im p<m$, the contour can be the positive real axis $C_{R0}$. 
    (b) for $\im p>m$, either $C_R$ or its deformation $C_R^{\prime}$ (blue dashed) can be the contour. 
    (c) for $\im p>m$, if $\re p >0$ approaches zero, the contour can be deformed into $C_L$, and $n_+=1$ between the two cut crossings. 
    (d) The same momentum as in (c), with $\ell_+$ reexpressed as a sum of two logarithms in Eq.~\eqref{eq:leftward_logs} and $C_L$ deformed into $C_L'$.}
    \label{fig:lse_log_contours}
\end{figure}
We start from considering a real on-shell $p>0$. Because of the $+i\epsilon$ prescription, an integration contour along the positive real axis passes below the pole $q=+ \sqrt{p^2 + i\epsilon}$ ($i\epsilon$ is suppressed for simplicity in the following), shown as $C_{R0}$ in Fig.~\ref{fig:lse_log_contours}~(a). 
We consider the case when $p$ is continued into the
first quadrant of the complex $p$-plane with $\re p >0$ and $\im p>0$.
We take the usual principal branch convention for logarithmic functions such that $\Log w$ has a cut along the negative real axis in the $w$ plane. Then the cut of each of the logarithmic functions in Eq.~\eqref{eq:contour_triangle} maps to two cuts in the complex $q$ plane:
\begin{align}
    \Log w_-:&\quad \re q =\re p,\quad |\im q-\im p|\ge m,\notag\\
    \Log w_+:&\quad \re q =-\re p,\quad |\im q+\im p|\ge m,
    \label{eq:triangle_principal_cuts}
\end{align} 
as shown by the dashed vertical lines in Fig.~\ref{fig:lse_log_contours}~(a).
In the plot, the labels $w_\pm \pm i \epsilon$, with $w_\pm <0$ on a cut, indicate the two edges of the cut where the principal values $\Log(w_\pm\pm i\epsilon)=\log|w_\pm|\pm i\pi$.

If $\im p>m$, the contour needs to be deformed to avoid crossing the logarithmic cut starting from $p-im$ into a different sheet of $\Log w_-$. For instance, we can deform the contour into $C_R$ with the path $0\to ih \to +\infty + ih$ $(\im p-m < h < \im p)$, as shown in Fig.~\ref{fig:lse_log_contours}~(b). Since the integration contour does not cross the logarithmic cuts, along the contour neither logarithmic argument crosses its own negative real axis, and thus they remain on the principal branches with $n_+ = n_-=0$.
Alternatively, the contour $C_R$ can be deformed into the indented contour $C_R'$~\cite{Dawid:2023jrj, Hu:2026gbf}, which can be evaluated as the integration over the principal logarithmic branches along the real axis of Eq.~\eqref{eq:contour_triangle} plus an additional piece along the two edges of the $p-im$ cut given by $2\pi i\int_{\re p}^{p-im}\!\frac{q\,dq}{p^2-q^2+i\epsilon}$ (the contribution around the infinitesimal arc around the branch point $p-im$  is not written out as it vanishes as the arc radius goes to zero).
The result will be incorrect if this piece is ignored.

Next, we consider the case with $p^2<0$ on the first Riemann sheet of on-shell energy with $\im p >m$. This is achieved with $\re p$ approaching $0^+$ and thus the logarithmic branch point $p-im$ is close to the imaginary $q$ axis.
To keep the integration contour away from this branch point such that the numerical calculation can be done easily, the contour can be deformed into $C_L$ shown in Fig.~\ref{fig:lse_log_contours}~(c).
However, in this case, the logarithmic argument $w_+$ crosses the negative real axis in its complex $w_+$ plane from above to below at $q = -\re p$ and then crosses back at $q = -\re p + i h$. During this process, the value of $n_+$ changes from 0 to 1 and then back to 0. 
Therefore, the result should be the integral of the principal logarithmic branches along $C_L$ plus $-2 \pi i \int_{C_S} \frac{q d q}{p^2-q^2+i \epsilon}$, with $C_S$ denoting the path $-\re p \rightarrow -r \rightarrow-r+i h \rightarrow -\re p+i h$.
In this case, one can avoid crossing the logarithmic branch cut by writing 
\begin{align}
    \ell_+(q)
    =\Log u_-
    +\Log u_+, \qquad \text{with } u_\pm = q+p \pm im.
    \label{eq:leftward_logs}
\end{align}
Then the negative $u_\pm$ real axes are mapped into horizontal cuts from $-p \mp im$ to the left on the complex $q$ plane, as shown in Fig.~\ref{fig:lse_log_contours}~(d).
For $\im p > m$, both cuts lie below the real axis and do not intersect $C_L$, which can then be deformed into $C_L'$ for easy evaluation.

\end{appendix}

\bibliography{ref}

@article{Du:2024gzw,
    author = "Du, Meng-Lin and Guo, Feng-Kun and Wu, Bing",
    title = "{Effective-Range Expansion with a Long-Range Force}",
    eprint = "2408.09375",
    archivePrefix = "arXiv",
    primaryClass = "hep-ph",
    doi = "10.1103/cwdt-dj6z",
    journal = "Phys. Rev. Lett.",
    volume = "135",
    number = "1",
    pages = "011903",
    year = "2025"
}

@article{Du:2025vkm,
    author = "Du, Meng-Lin and Guo, Feng-Kun and Wu, Bing",
    title = "{Effective range expansion with the left-hand cut and its application to the $T_{cc}(3875)$}",
    eprint = "2502.19774",
    archivePrefix = "arXiv",
    primaryClass = "hep-ph",
    doi = "10.22323/1.479.0062",
    journal = "PoS",
    volume = "CD2024",
    pages = "062",
    year = "2026"
}

@article{Wang:2026ups,
    author = "Wang, Wen-Jia and Wu, Bing and Du, Meng-Lin and Guo, Feng-Kun",
    title = "{Effective range expansion with the left-hand cut: Higher order improvements}",
    eprint = "2601.04989",
    archivePrefix = "arXiv",
    primaryClass = "hep-ph",
    doi = "10.1103/tdyn-422c",
    journal = "Phys. Rev. D",
    volume = "114",
    number = "1",
    pages = "014003",
    year = "2026"
}

@article{Chew:1960iv,
	author = "Chew, Geoffrey F. and Mandelstam, Stanley",
	title = "{Theory of low-energy pion pion interactions}",
	doi = "10.1103/PhysRev.119.467",
	journal = "Phys. Rev.",
	volume = "119",
	pages = "467--477",
	year = "1960"
}

@book{Oller:2019rej,
    author = "Oller, Jos\'e Antonio",
    title = "{A Brief Introduction to Dispersion Relations}",
    doi = "10.1007/978-3-030-13582-9",
    isbn = "978-3-030-13581-2, 978-3-030-13582-9",
    publisher = "Springer",
    series = "SpringerBriefs in Physics",
    year = "2019"
}

@article{Jing:2025qmi,
    author = "Jing, Hao-Jie and Cao, Xiong-Hui and Guo, Feng-Kun",
    title = "{Discontinuity calculus and applications to two-body coupled-channel scattering}",
    eprint = "2507.06175",
    archivePrefix = "arXiv",
    primaryClass = "hep-ph",
    doi = "10.15302/frontphys.2026.056201",
    journal = "Front. Phys. (Beijing)",
    volume = "21",
    number = "5",
    pages = "056201",
    year = "2026"
}

@article{Du:2021zzh,
    author = "Du, Meng-Lin and Baru, Vadim and Dong, Xiang-Kun and Filin, Arseniy and Guo, Feng-Kun and Hanhart, Christoph and Nefediev, Alexey and Nieves, Juan and Wang, Qian",
    title = "{Coupled-channel approach to $T_{cc}^+$ including three-body effects}",
    eprint = "2110.13765",
    archivePrefix = "arXiv",
    primaryClass = "hep-ph",
    doi = "10.1103/PhysRevD.105.014024",
    journal = "Phys. Rev. D",
    volume = "105",
    number = "1",
    pages = "014024",
    year = "2022"
}

@article{Cutkosky:1960sp,
    author = "Cutkosky, R. E.",
    title = "{Singularities and discontinuities of Feynman amplitudes}",
    doi = "10.1063/1.1703676",
    journal = "J. Math. Phys.",
    volume = "1",
    pages = "429--433",
    year = "1960"
}

@book{Eden:1966dnq,
    author = "Eden, Richard John and Landshoff, Peter V. and Olive, David I. and Polkinghorne, John Charlton",
    title = "{The analytic S-matrix}",
    isbn = "978-0-521-04869-9",
    publisher = "Cambridge Univ. Press",
    address = "Cambridge",
    year = "1966"
}

@book{Martin:1970hmp,
    author = "Martin, A. D. and Spearman, T. D.",
    title = "{Elementary Particle Theory}",
    isbn = "978-0-7204-0157-8",
    publisher = "North-Holland",
    address = "Amsterdam",
    year = "1970"
}

@article{Wu:2023uva,
    author = "Wu, Bing and Cao, Xiong-Hui and Dong, Xiang-Kun and Guo, Feng-Kun",
    title = "{\ensuremath{\sigma} exchange in the one-boson exchange model involving the ground state octet baryons}",
    eprint = "2312.01013",
    archivePrefix = "arXiv",
    primaryClass = "hep-ph",
    doi = "10.1103/PhysRevD.109.034026",
    journal = "Phys. Rev. D",
    volume = "109",
    number = "3",
    pages = "034026",
    year = "2024"
}

@article{Wu:2024xwy,
    author = "Wu, Bing and Dong, Xiang-Kun and Du, Meng-Lin and Guo, Feng-Kun and Zou, Bing-Song",
    title = "{Deciphering the mechanism of $J/\psi$-nucleon scattering}",
    eprint = "2410.19526",
    archivePrefix = "arXiv",
    primaryClass = "hep-ph",
    doi = "10.1016/j.fmre.2025.07.005",
    journal = "Fund. Res.",
    volume = "5",
    pages = "2530--2536",
    year = "2025"
}

@article{Bethe:1949yr,
	author = "Bethe, H. A.",
	title = "{Theory of the Effective Range in Nuclear Scattering}",
	doi = "10.1103/PhysRev.76.38",
	journal = "Phys. Rev.",
	volume = "76",
	pages = "38--50",
	year = "1949"
}

@article{Blatt:1949zz,
	author = "Blatt, John M. and Jackson, J. David",
	title = "{On the Interpretation of Neutron-Proton Scattering Data by the Schwinger Variational Method}",
	doi = "10.1103/PhysRev.76.18",
	journal = "Phys. Rev.",
	volume = "76",
	pages = "18--37",
	year = "1949"
}

@article{Liu:2026xrk,
    author = "Liu, Bo-Yang and Wu, Bing and Fu, Ji-Wei and Du, Meng-Lin and Guo, Feng-Kun and Mei{\ss}ner, Ulf-G.",
    title = "{Extraction of the pion-nucleon coupling constant using the effective-range expansion with the left-hand cut}",
    eprint = "2602.03115",
    archivePrefix = "arXiv",
    primaryClass = "hep-ph",
    doi = "10.1103/184t-bpbg",
    journal = "Phys. Rev. D",
    volume = "114",
    number = "7",
    pages = "074030",
    year = "2026"
}

@article{Dong:2020hxe,
    author = "Dong, Xiang-Kun and Guo, Feng-Kun and Zou, Bing-Song",
    title = "{Explaining the Many Threshold Structures in the Heavy-Quark Hadron Spectrum}",
    eprint = "2011.14517",
    archivePrefix = "arXiv",
    primaryClass = "hep-ph",
    doi = "10.1103/PhysRevLett.126.152001",
    journal = "Phys. Rev. Lett.",
    volume = "126",
    number = "15",
    pages = "152001",
    year = "2021"
}

@article{Braaten:2005jj,
    author = "Braaten, Eric and Kusunoki, Masaoki",
    title = "{Factorization in the production and decay of the $X(3872)$}",
    eprint = "hep-ph/0506087",
    archivePrefix = "arXiv",
    doi = "10.1103/PhysRevD.72.014012",
    journal = "Phys. Rev. D",
    volume = "72",
    pages = "014012",
    year = "2005"
}

@article{Cohen:2004kf,
    author = "Cohen, Thomas D. and Gelman, Boris A. and van Kolck, U.",
    title = "{An Effective field theory for coupled channel scattering}",
    eprint = "nucl-th/0402054",
    archivePrefix = "arXiv",
    doi = "10.1016/j.physletb.2004.03.020",
    journal = "Phys. Lett. B",
    volume = "588",
    pages = "57--66",
    year = "2004"
}

@article{Guo:2013rpa,
    author = "Guo, Zhi-Hui and Oller, J. A. and R{\'\i}os, G.",
    title = "{Nucleon-Nucleon scattering from the dispersive N/D method: next-to-leading order study}",
    eprint = "1305.5790",
    archivePrefix = "arXiv",
    primaryClass = "nucl-th",
    doi = "10.1103/PhysRevC.89.014002",
    journal = "Phys. Rev. C",
    volume = "89",
    number = "1",
    pages = "014002",
    year = "2014"
}

@article{Hoferichter:2012wf,
    author = "Hoferichter, M. and Ditsche, C. and Kubis, B. and Meissner, U. G.",
    title = "{Dispersive analysis of the scalar form factor of the nucleon}",
    eprint = "1204.6251",
    archivePrefix = "arXiv",
    primaryClass = "hep-ph",
    doi = "10.1007/JHEP06(2012)063",
    journal = "JHEP",
    volume = "06",
    pages = "063",
    year = "2012"
}

@article{Hanhart:2011jz,
    author = "Hanhart, C. and Kalashnikova, Yu. S. and Nefediev, A. V.",
    title = "{Interplay of quark and meson degrees of freedom in a near-threshold resonance: multi-channel case}",
    eprint = "1106.1185",
    archivePrefix = "arXiv",
    primaryClass = "hep-ph",
    doi = "10.1140/epja/i2011-11101-9",
    journal = "Eur. Phys. J. A",
    volume = "47",
    pages = "101--110",
    year = "2011"
}

@article{Bjorken:1960zz,
    author = "Bjorken, J. D.",
    title = "{Construction of Coupled Scattering and Production Amplitudes Satisfying Analyticity and Unitarity}",
    doi = "10.1103/PhysRevLett.4.473",
    journal = "Phys. Rev. Lett.",
    volume = "4",
    pages = "473--474",
    year = "1960"
}

@article{Kaplan:1998we,
    title = {Two Nucleon Systems from Effective Field Theory},
    author = {Kaplan, David B. and Savage, Martin J. and Wise, Mark B.},
    year = {1998},
    journal = {Nucl. Phys. B},
    volume = {534},
    eprint = {nucl-th/9802075},
    pages = {329--355},
    doi = {10.1016/S0550-3213(98)00440-4},
    archiveprefix = {arXiv}
}

@article{Du:2023hlu,
    title = {Role of {{Left-Hand Cut Contributions}} on {{Pole Extractions}} from {{Lattice Data}}: {{Case Study}} for {{Tcc}}(3875){$^{+}$}},
    author = {Du, Meng-Lin and Filin, Arseniy and Baru, Vadim and Dong, Xiang-Kun and Epelbaum, Evgeny and Guo, Feng-Kun and Hanhart, Christoph and Nefediev, Alexey and Nieves, Juan and Wang, Qian},
    year = {2023},
    journal = {Phys. Rev. Lett.},
    volume = {131},
    eprint = {2303.09441},
    primaryclass = {hep-ph},
    pages = {131903},
    doi = {10.1103/PhysRevLett.131.131903},
    archiveprefix = {arXiv}
}

@article{Zhang:2023wdz,
    title = {Neutron {{Scattering}} off {{One-Neutron Halo Nuclei}} in {{Halo Effective Field Theory}}},
    author = {Zhang, Xu and Fu, Hai-Long and Guo, Feng-Kun and Hammer, Hans-Werner},
    year = {2023},
    journal = {Phys. Rev. C},
    volume = {108},
    eprint = {2308.12815},
    primaryclass = {nucl-th},
    pages = {044304},
    doi = {10.1103/PhysRevC.108.044304},
    archiveprefix = {arXiv}
}

@article{ParticleDataGroup:2026mpi,
    title = {Review of {{Particle Physics}}},
    author = {Takahashi, F. and others},
    year = {2026},
    journal = {Int. J. Mod. Phys. A},
    volume = {41},
    pages = {2630011},
    doi = {10.1142/s0217751x26300115},
    collaboration = {Particle Data Group}
}

@article{Ji:2025hjw,
    author = "Ji, Teng and Dong, Xiang-Kun and Guo, Feng-Kun and Hanhart, Christoph and Mei{\ss}ner, Ulf-G.",
    title = "{Precise determination of the properties of $X(3872)$ and of its isovector partner $W_{c1}$}",
    eprint = "2502.04458",
    archivePrefix = "arXiv",
    primaryClass = "hep-ph",
    month = "2",
    year = "2025",
    journal = "Sci. Bull.",
    pages = "in print"
}

@article{BESIII:2023hml,
    author = "Ablikim, Medina and others",
    collaboration = "BESIII",
    title = "{Coupled-Channel Analysis of the ${\chi}_{c1}(3872)$ Line Shape with BESIII Data}",
    eprint = "2309.01502",
    archivePrefix = "arXiv",
    primaryClass = "hep-ex",
    doi = "10.1103/PhysRevLett.132.151903",
    journal = "Phys. Rev. Lett.",
    volume = "132",
    number = "15",
    pages = "151903",
    year = "2024"
}

@article{LHCb:2020fvo,
    author = "Aaij, Roel and others",
    collaboration = "LHCb",
    title = "{Study of the $\psi_2(3823)$ and $\chi_{c1}(3872)$ states in $B^+ \rightarrow \left( J\psi\pi^+\pi^-\right)K^+$ decays}",
    eprint = "2005.13422",
    archivePrefix = "arXiv",
    primaryClass = "hep-ex",
    reportNumber = "CERN-EP-2020-071, LHCb-PAPER-2020-009",
    doi = "10.1007/JHEP08(2020)123",
    journal = "JHEP",
    volume = "08",
    pages = "123",
    year = "2020"
}

@article{LHCb:2022jez,
    author = "Aaij, Roel and others",
    collaboration = "LHCb",
    title = "{Observation of sizeable $\omega$ contribution to $\chi_{c 1}(3872) \rightarrow \pi^{+} \pi^{-} J / \psi$ decays}",
    eprint = "2204.12597",
    archivePrefix = "arXiv",
    primaryClass = "hep-ex",
    reportNumber = "LHCb-PAPER-2021-045, CERN-EP-2022-049",
    doi = "10.1103/PhysRevD.108.L011103",
    journal = "Phys. Rev. D",
    volume = "108",
    number = "1",
    pages = "L011103",
    year = "2023"
}

@article{Ross:1960mert,
    author = "Ross, Marc H. and Shaw, Gordon L.",
    title = "{Scattering length and effective range theory for multichannel processes}",
    doi = "10.1016/0003-4916(60)90040-3",
    journal = "Annals Phys.",
    volume = "9",
    number = "3",
    pages = "391--415",
    year = "1960"
}

@article{Ross:1961jlg,
    author = "Ross, Marc H. and Shaw, Gordon L.",
    title = "{Multichannel effective range theory}",
    doi = "10.1016/0003-4916(61)90078-1",
    journal = "Annals Phys.",
    volume = "13",
    number = "2",
    pages = "147--186",
    year = "1961"
}

@article{CLQCD:2019npr,
    author = "Chen, Ting and Chen, Ying and Gong, Ming and Liu, Chuan and Liu, Liuming and Liu, Yu-Bin and Liu, Zhaofeng and Ma, Jian-Ping and Werner, Markus and Zhang, Jian-Bo",
    collaboration = "CLQCD",
    title = "{A coupled-channel lattice study on the resonance-like structure $Z_c(3900)$}",
    eprint = "1907.03371",
    archivePrefix = "arXiv",
    primaryClass = "hep-lat",
    doi = "10.1088/1674-1137/43/10/103103",
    journal = "Chin. Phys. C",
    volume = "43",
    number = "10",
    pages = "103103",
    year = "2019"
}

@article{Entem:2016ipb,
    author = {Entem, D. R. and Oller, J. A.},
    title = {{The N/D method with non-perturbative left-hand-cut discontinuity and the $^1S_0$ $NN$ partial wave}},
    eprint = {1610.01040},
    archivePrefix = {arXiv},
    primaryClass = {nucl-th},
    doi = {10.1016/j.physletb.2017.09.012},
    journal = {Phys. Lett. B},
    volume = {773},
    pages = {498--504},
    year = {2017}
}

@article{Nieves:2003uu,
    author = {Nieves, J.},
    title = {{Renormalization of the one pion exchange NN interaction in presence of derivative contact interactions}},
    eprint = {nucl-th/0301080},
    archivePrefix = {arXiv},
    doi = {10.1016/j.physletb.2003.05.009},
    journal = {Phys. Lett. B},
    volume = {568},
    pages = {109--117},
    year = {2003}
}

@article{Entem:2025siq,
    author = {Entem, David R. and Nieves, Juan and Oller, Jose Antonio},
    title = {{Contact potentials in the presence of a regular finite-range interaction using dimensional regularization and the N/D method}},
    eprint = {2506.01461},
    archivePrefix = {arXiv},
    primaryClass = {nucl-th},
    doi = {10.1103/n2x6-16rc},
    journal = {Phys. Rev. D},
    volume = {112},
    number = {9},
    pages = {096007},
    year = {2025}
}

@article{Oller:2018zts,
    author = {Oller, J. A. and Entem, D. R.},
    title = {{The exact discontinuity of a partial wave along the left-hand cut and the exact $N/D$ method in non-relativistic scattering}},
    eprint = {1810.12242},
    archivePrefix = {arXiv},
    primaryClass = {hep-ph},
    doi = {10.1016/j.aop.2019.167965},
    journal = {Annals Phys.},
    volume = {411},
    pages = {167965},
    year = {2019}
}

@article{Landau:1959fi,
    author = {Landau, L. D.},
    editor = {ter Haar, D.},
    title = {{On the Analytic Properties of Vertex Parts in Quantum Field Theory}},
    doi = {10.1016/B978-0-08-010586-4.50103-6},
    journal = {Zh. Eksp. Teor. Fiz.},
    volume = {37},
    number = {1},
    pages = {62--70},
    year = {1960}
}

@article{Bayar:2016ftu,
    author = {Bayar, Melahat and Aceti, Francesca and Guo, Feng-Kun and Oset, Eulogio},
    title = {{A Discussion on Triangle Singularities in the $\Lambda_b \to J/\psi K^{-} p$ Reaction}},
    eprint = {1609.04133},
    archivePrefix = {arXiv},
    primaryClass = {hep-ph},
    doi = {10.1103/PhysRevD.94.074039},
    journal = {Phys. Rev. D},
    volume = {94},
    number = {7},
    pages = {074039},
    year = {2016}
}

@article{Dawid:2023jrj,
    author = {Dawid, Sebastian M. and Islam, Md Habib E. and Brice{\~n}o, Ra{\'u}l A.},
    title = {{Analytic continuation of the relativistic three-particle scattering amplitudes}},
    eprint = {2303.04394},
    archivePrefix = {arXiv},
    primaryClass = {nucl-th},
    doi = {10.1103/PhysRevD.108.034016},
    journal = {Phys. Rev. D},
    volume = {108},
    number = {3},
    pages = {034016},
    year = {2023}
}

@article{Hu:2026gbf,
    author = {Hu, Tao-Ran and Fu, Hai-Long and Guo, Feng-Kun and Mei{\ss}ner, Ulf-G. and Zhang, Xu},
    title = {{Three-body unitary determination of the $f_1(1285)$ and $f_1(1420)$ pole positions}},
    journal = {Commun. Theor. Phys.},
    pages = {in print},
    eprint = {2606.19040},
    archivePrefix = {arXiv},
    primaryClass = {hep-ph},
    month = {6},
    year = {2026}
}

@article{Blatt:1952zz,
    author = "Blatt, John M. and Biedenharn, L. C.",
    title = "{The Angular Distribution of Scattering and Reaction Cross Sections}",
    doi = "10.1103/RevModPhys.24.258",
    journal = "Rev. Mod. Phys.",
    volume = "24",
    pages = "258--272",
    year = "1952"
}

@article{Stapp:1956mz,
    author = "Stapp, H. P. and Ypsilantis, T. J. and Metropolis, N.",
    title = "{Phase shift analysis of 310-MeV proton proton scattering experiments}",
    doi = "10.1103/PhysRev.105.302",
    journal = "Phys. Rev.",
    volume = "105",
    pages = "302--310",
    year = "1957"
}

\end{document}